\documentclass[11pt,a4paper]{article}
\usepackage[inner=2.5cm,outer=2.5cm, top=2.5cm, bottom=2.5cm]{geometry}
\usepackage{authblk}
\usepackage[english]{babel}
\usepackage[utf8]{inputenc}
\usepackage{graphicx}
\usepackage{amsmath}
\usepackage{float}
\usepackage{siunitx}
\usepackage[normalem]{ulem}

\usepackage[svgnames]{xcolor}
\usepackage[hidelinks]{hyperref}
\usepackage[capitalise,nameinlink]{cleveref}
\usepackage{mathrsfs}
\usepackage{textgreek}
\usepackage[caption = false]{subfig}

\title{Development of a Prediction Model for Indoor Rolling Noise}

\author[1]{M. Edwards\thanks{matthew.edwards@matelys.com}}
\author[1]{F. Chevillotte}
\author[1]{F.-X. Bécot}
\author[1]{L. Jaouen}
\author[2]{N. Totaro}

\affil[1]{Matelys Research Lab, 7 Rue des Maraîchers, Bât B, 69120 Vaulx-en-Velin, France}
\affil[2]{INSA---Lyon, 20 Avenue Albert Einstein, 69621 Villeurbanne cedex, France}

\date{May 5 2021}

\begin{document}

\maketitle

\begin{abstract}
This work presents a prediction model for rolling noise in multi-story buildings, such as that generated by a rolling delivery trolley. Until now, mechanical excitation in multi-story buildings has been limited to impact sources such as the tapping machine. Rolling noise models have been limited to outdoor sources such as trains and automotive vehicles. The model presented here is able to represent the physical phenomena unique to indoor rolling noise, taking into account influencing factors such as the roughness of the wheel and the floor, the material and geometric properties of the wheel and the floor, the rolling velocity of the trolley, and the load on the trolley. The model may be used as a tool to investigate how different flooring systems (including multi-layer systems) respond to rolling excitation, for the purpose of developing multi-story building solutions which are better equipped to combat this kind of noise source.
\end{abstract}

\section{Introduction} \label{sec:introduction}
Structure-borne noise insulation is a field of building acoustics which has seen a large amount of research in recent decades. By far and away, the most common method of evaluating a building's performance in mitigating structure-borne noise is linked to impact noise. This is done primarily through the use of a tapping machine \cite{ISO10140320102010}, though other methods such as the rubber ball \cite{ISO10140520102010} (see appendix F), walking noise \cite{BS1620520132018}, and even rainfall noise \cite{ISO10140520102010} are sometimes used. Prediction of upward and lateral radiation from airborne and structure-borne excitation in 
buildings exists as well \cite{ISO12354220172017}. 
However, other sources exist beyond these, such as rolling noise. 

\begin{figure}[ht!]
	\centering
	\includegraphics[width=0.4\textwidth]{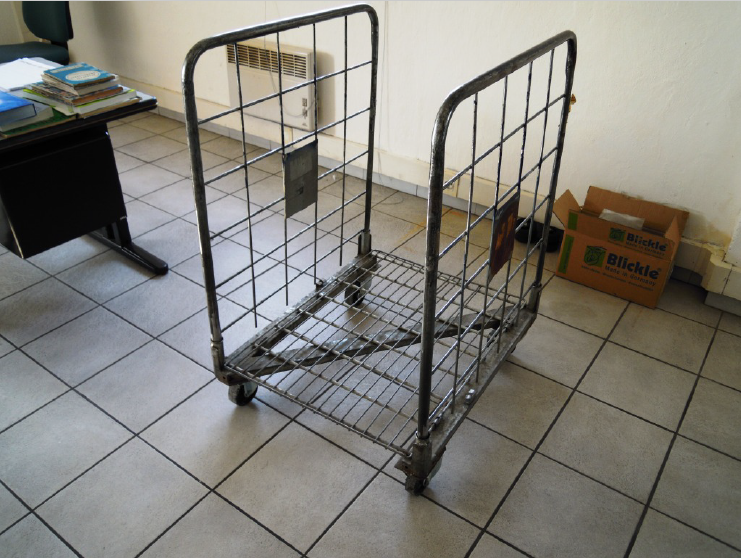}
	\caption{A typical rolling delivery trolley.}
	\label{fig:trolley_cart}
\end{figure}

Indoor rolling noise can take many forms: rolling office chairs, children's toys, and suitcases being a few examples of such structure-borne excitation sources. The most annoying, however, are possibly rolling trolleys or delivery carts: an example of which is shown in \cref{fig:trolley_cart}. Used to conduct deliveries and move merchandise in commercial spaces, these devices generate a constant, low frequency noise which propagates easily through the structure, disturbing the inhabitants of the apartments which often exist above these commercial spaces in urban areas. This noise is generated by the excitation of the wheels and the floor due to the relative small-scale surface roughness between them \cite{remingtonWheelRailRolling1988}.

\begin{figure}[ht!]
	\centering
	\includegraphics{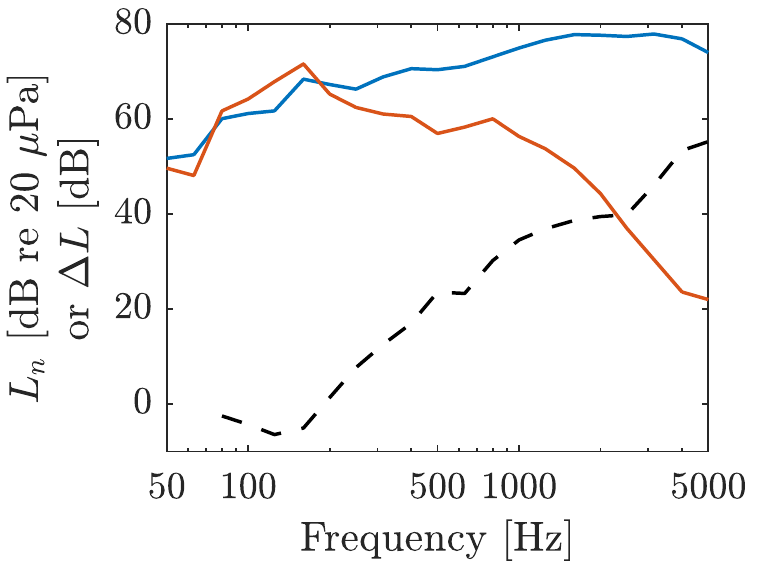}
	\caption{Comparison of the spectra of tapping noise and rolling noise of a classical concrete floor (140~mm), as well as the attenuation of a classical floating floor (140~mm concrete slab + a decoupling layer + 40~mm screed). \cite{edwardsRollingNoiseModeling2018}. \mbox{\protect\includegraphics{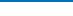} $L_n$: tapping} machine, \mbox{\protect\includegraphics{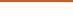} $L_n$: rolling} trolley, \mbox{\protect\includegraphics{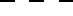} $\Delta L$: attenuation} of a classical floating floor.}
	\label{fig:spectra_comp}
\end{figure}

\cref{fig:spectra_comp} shows the acoustic excitation due to a standard tapping machine and a typical rolling trolley on a classical concrete floor, in addition to the attenuation of a classical floating floor. The spectrum of impact noise is quite different than that of rolling noise, having the majority of its acoustic energy in the low frequency range. Furthermore, the attenuation of a classical floating floor is inverse to said rolling noise profile. As such, it does not provide sufficient insulation on its own without additional treatment.

The difficulty in combating this type of structure-borne noise is that, because its sound signature is starkly different than that of impact noise, the solutions developed to help against impact noise can be sometimes ineffective against rolling noise. Additionally, impact noise modeling techniques cannot accurately predict the level of rolling noise generated in the same environment. For this reason, a unique method of modeling and predicting indoor rolling noise is needed.

Rolling noise has indeed been thoroughly investigated for decades, but uniquely in the context of vehicle tire/road and train wheel/rail contact. Models which aim at predicting these types of noise exist (see for example \cite{wullensThreeDimensionalContactModel2004,anderssonTimeDomainContact2008,pinningtonTyreRoadContact2013} for tire and \cite{remingtonEstimationWheelRail1996,thompsonExperimentalValidationTWINS1996a,thompsonExperimentalValidationTWINS1996,lundbergNonlinearStateDependentModel2015} for train), but they cannot be easily adapted to work for indoor rolling noise due to differences in the underlying phenomena which cause such noise (mainly the propagation of the sound to the surrounding environment, which is quite different between roads, train tracks, and indoor buildings).

With train wheel/rail contact, the rail is often modeled using techniques such as an infinite beam \cite{remingtonWheelRailNoise1976a,wuDoubleTimoshenkoBeam1999,thompsonExperimentalValidationTWINS1996a}, moving Green's functions \cite{maziluGreenFunctionsAnalysis2007,nordborgWheelRailNoise2002}, and finite element (FE) modeling \cite{nielsenVerticalDynamicInteraction1995}, using either discrete or continuous supports. 

With vehicle tire/road contact, the road is often given less focus, considering its more simple construction as a flat plane. Some models (e.g. \cite{becotSoundRadiationTyres2000,kleinTyreRoadNoise2008}) incorporate the presence of the road in order to calculate its influence on the sound radiation of the tire, though do not treat it as a vibrating plate itself for the purpose of sound radiation. This is often done to model phenomena such as the horn effect between the tire and the road (described in \cite{wullensThreeDimensionalContactModel2004}), which is known to play a large role in vehicle tire/road contact noise. Priority is instead given to the modeling of the tire itself, whose vibration contributes significantly to the overall radiated sound. Various models may treat the tire as either a vibrating cylindrical ring \cite{kindtDevelopmentValidationThreeDimensional2009}, an unfolded vibrating plate \cite{larssonHighFrequencyThreeDimensionalTyre2002}, or even a waveguide \cite{hoeverModelInvestigatingInfluence2015}.

Conversely, trolley wheel/floor contact has several differences which do not allow for the direct application of these previously established techniques. The environment in which the sound propagates is starkly different, as the primary transfer path is structure-borne (vertically through the floor) rather than airborne (horizontally away from the rail or road). The forces at play are much smaller (due to the significantly lower mass of a trolley than that of a car or train), as is the overall geometry. Here the floor acts as a vibrating plate, being excited by a continuous (and moving) injected force. In train wheel/rail contact, the two surfaces are made of the same (or similar) metallic material, and are extremely stiff. In vehicle tire/road contact, the tire is significantly softer than the road, and this high difference in elasticity plays a role in how their contact is modeled (through the use of an envelopment procedure \cite{beckenbauerTyreRoadNoise2008,kleinTyreRoadNoise2008}, for example). In trolley wheel/floor contact, neither of these two situations exist. Dozens of different materials are used to make indoor floors and trolley wheels: varying greatly in not only elasticity, but roughness as well. An indoor rolling noise model should be capable of handling such a wide range of materials.

Previously, we developed a simple model for predicting rolling noise in an indoor environment \cite{chevillotteRollingNoiseModel2015,edwardsRollingNoiseModeling2018}. This model calculates the contact force between the wheel and the floor in the time domain, and uses it as the injected force into the floor to predict the resulting radiated sound power. Here we present an improved version of this model, which greatly expands in scope and application the capabilities of predicting indoor rolling noise. The improved model is capable of accounting for floors of various types and layered constructions, as well as discrete irregularities (beyond small-scale surface roughness) such as wheel flat spots and floor joints. In the sections that follow, the model is presented, along with a comparison with experimental test results.

\section{Model development}\label{sec:model}
\begin{figure}[!htbp]
	\centering
	\includegraphics{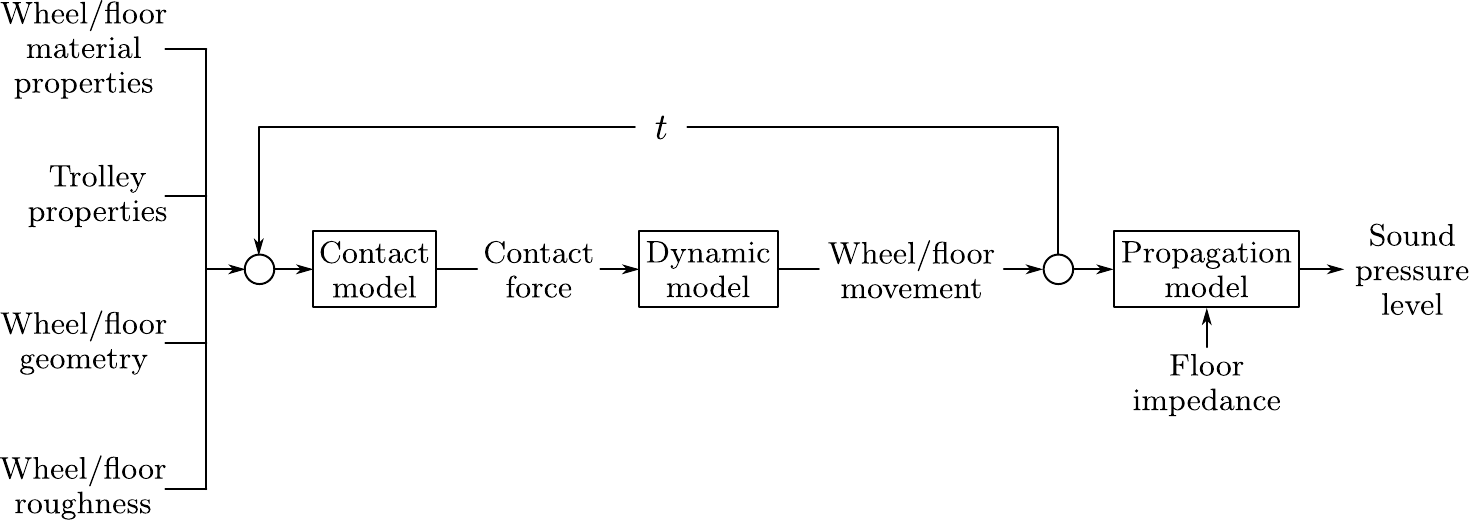}
	\caption{General outline of the rolling noise model.}
	\label{fig:block_chart_hor}
\end{figure}

\cref{fig:block_chart_hor} shows the general outline of the rolling noise model. Operating in the time domain and in three spatial dimensions, the model uses the influencing properties of the wheel and floor to calculate the contact force between the two bodies for each discretized moment in time throughout the rolling event. These include the wheel and floor roughness profiles, the wheel and floor material properties (Young's modulus and Poisson's ratio), the speed of the trolley, the mass of the trolley (plus added load), and the geometric profiles of the wheel and floor. Such profiles include not only the curvature of the wheel, but also any discrete irregularities such as wheel flats or floor joints.

The contact force at each discretized moment in time is used as input into the dynamic model, where the resulting movement of the wheel and floor are calculated in response to the excitation injected force. This influences what the exact roughness profile will be for the following time step, as the wheel and floor continuously move in relation to one another: influencing how much (or how little) contact exists between them. Once the contact force has been calculated for each time step in the rolling event, the model is converted to the frequency domain, where the blocked force spectra is used in conjunction with the wheel and floor impedances to convert the blocked force to an injected force. From there the resulting radiated sound power may be calculated in the reception room.

\subsection{Contact model} \label{sec:contact}
\begin{figure}[!htbp]
	\centering
	\includegraphics{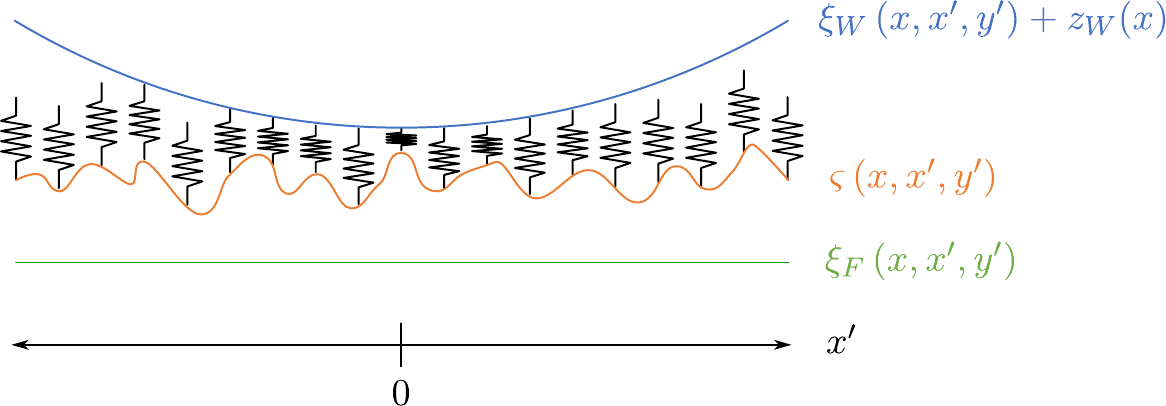}
	\caption{Contact model used for estimating the contact stresses between the wheel and the floor}
	\label{fig:contact_model}
\end{figure}

In order to estimate the contact force for each time step, a Winkler bedding is used. This has been used in other rolling noise models (for example \cite{thompsonRelationshipWheelRail1996,pieringerTimeDomainModel2007}), and has proven to be a good estimation when compared to the more physically accurate Boussinesq method \cite{boussinesqApplicationPotentielsEtude1885}. \cref{fig:contact_model} shows a diagram of the contact model. A bed of independent, locally reacting springs is placed between the wheel and the floor, and their deflection used to estimate the stress profile in the area of contact for each discretized moment in time.

The geometry of this interaction is described by $u_R$, the interpenetration of the two surfaces
\begin{equation} \label{eq:uprimer}
u_R(x,x',y') = u_0 - z_W(x) + \varsigma(x,x',y') - \xi_W(x,x',y') + \xi_F(x,x')
\end{equation}
Here $u_0$ is the static Hertzian interpenetration, $z_W$ is the vertical position of the wheel center, $\varsigma$ is the relative roughness profile between the wheel $\varsigma_W$ and the floor $\varsigma_F$
\begin{equation} \label{eq:uR}
\varsigma(x,x',y') = \varsigma_F(x,x',y') - \varsigma_W(x,x',y')
\end{equation}
and $\xi_W$ and $\xi_F$ are the geometric profiles of the wheel and the floor respectively.

The static Hertzian interpenetration is the interpenetration of the two bodies in the absence of roughness. For a general ellipsoidal wheel (i.e. one with both a normal and a transverse radius of curvature), this relationship is described by \cref{eq:u0_ell}
\begin{equation} \label{eq:u0_ell}
u_0 = \frac{u^*\Sigma\rho}{2}\left(\frac{3Q}{2E'\Sigma\rho}\right)^{2/3}
\end{equation}
where $Q$ is the effective mass $M$ expressed as a load in Newtons and $\Sigma\rho$ is the curvature sum for contact between an ellipsoid and an elastic half space.
\begin{equation} \label{eq:rho}
\Sigma\rho=\frac{1}{r_x}+\frac{1}{r_y}
\end{equation}
Here $r_x$ and $r_y$ are the radii of curvature in the $x$ and $y$ directions.

$E'$ is the apparent Young's modulus between the wheel and the floor
\begin{equation} \label{eq:Eapp}
E' = \left(\frac{1-\nu_W^2}{E_W}+\frac{1-\nu_F^2}{E_F}\right)^{-1}
\end{equation}
where $E$ and $\nu$ are the Young's modulus and Poisson's ratio of the wheel and floor (denoted by the appropriate subscripts).

The dimensionless quantity $u^*$ is a function of only the body geometries and material properties. A thorough explanation of how this is derived is given by Harris in \cite{harrisRollingBearingAnalysis2007} (see chapter 6).

For the case of a cylindrical wheel with no transverse radius, \cref{eq:u0_ell} simplifies to
\begin{equation} \label{eq:u0_cyl}
u_0 = \frac{4Q}{E'\pi w}
\end{equation}
where $w$ is the width of the wheel in the $y$ direction. In the absence of roughness, the contact area between a cylindrical wheel and flat floor is a rectangle of length $2a$ in the longitudinal direction and width $2b$ in the lateral direction (equal to the wheel width $w$). For an ellipsoidal wheel, the contact area becomes an ellipse with dimensions $2a$ in the longitudinal direction and $2b$ in the lateral direction.

For a given moment in time $t$, the wheel center is at a position $x$ in the global coordinate system. The two quantities are related via the speed of the trolley $v$ such that $x = vt$. The local coordinate system is positioned such that $x'(0) = x$. Thus, the origin of the local system follows the wheel center in the $x$ dimension.

The contact springs in the Winkler bedding are set to deform with the square root of their deflection, in order to ensure agreement with Hertzian contact theory \cite{remingtonEstimationWheelRail1996}. This is seen in the equation used to estimate the stress profile in the area of contact
\begin{equation} \label{eq:sigmar}
\sigma_R(x,x',y') = \begin{cases} 
\sigma_0\sqrt{\frac{u_R(x,x',y')}{u_0}} & u_R(x,x',y') > 0 \\
0 & u_R(x,x',y') \leq 0 
\end{cases}
\end{equation}
Any points in the discretized calculation area which are found to have negative interpenetration are assumed to be out of contact, and the stresses at those points are set to zero.

Modifying the deflection of the contact springs also requires one to modify the radii of curvature used in defining the wheel profile $\xi_W$. The radii must be reduced in order to ensure continued agreement with Hertzian contact mechanics, otherwise the model would estimate an area of contact which is larger than reality. This is done with the following relations
\begin{equation} \label{eq:rxprime_ell}
r'_x = \frac{{a^*}^2}{u^*\Sigma\rho}
\end{equation}
\begin{equation} \label{eq:ryprime_ell}
r'_y = \frac{{b^*}^2}{u^*\Sigma\rho}
\end{equation}
where $a^*$ and $b^*$ are further dimensionless quantities defined in \cite{harrisRollingBearingAnalysis2007}. For a cylindrical wheel, \cref{eq:rxprime_ell} reduces to
\begin{equation} \label{eq:rxprime_cyl}
r'_x = \frac{r_x}{2}
\end{equation}

For an ellipsoidal wheel, the geometric dimensions $a$ and $b$ are
\begin{equation} \label{eq:a}
a=\sqrt{2r'_x u_0}
\end{equation}
\begin{equation} \label{eq:b}
b=\sqrt{2r'_y u_0}
\end{equation}
where $r'_x$ and $r'_y$ are given in \cref{eq:rxprime_ell,eq:ryprime_ell}. For a cylindrical wheel, the formulation of $r'_x$ given in \cref{eq:rxprime_cyl} is used, and $b$ is instead defined by half the wheel width $w/2$.

For a given $x$, the roughness profile $\varsigma$, wheel profile $\xi_W$ and floor profile $\xi_F$ are re-defined for the local coordinate system. Typically, the wheel profile will be simply defined by it's radii of curvature.
\begin{equation} \label{eq:xi_Wsmooth}
\xi_W(x,x',y') = \frac{{x'}^2}{2r'_x} + \frac{{y'}^2}{2r'_y}
\end{equation}
The floor profile will typically be zero for all $x$. However they may differ in the presence of discrete irregularities, and the profiles will need to be updated at each $x$ as a flat spot or floor joint moves through the vicinity of contact. The process for defining these more complex profiles is given in \cref{sec:discrete_irregularities}.

For a cylindrical wheel rolling on a flat floor, and in the absence of roughness, the interpenetration profile would be
\begin{equation} \label{eq:uR_exp}
u_R(x,x') = u_0 - \frac{{x'}^2}{2r'_x}
\end{equation}

Once the stress profile is known, it may then be integrated across the entire contact area to obtain the full contact force due to roughness,
\begin{equation} \label{eq:Fr}
F_R(x) = \int\int\sigma_R(x,x',y')\textrm{d}x'\textrm{d}y'
\end{equation}
Care must be taken to ensure the integration is carried out over a sufficiently large area so as to encompass the entire area of contact, which will not be exactly equal to $2a\times2b$ in the presence of roughness. When performing the integration numerically, the presence of any nonzero values on the edges of the stress matrix prior to integration indicates that the limits are too small.

\section{Dynamic model} \label{sec:dynamic_model}
\begin{figure}[!htbp]
	\centering
	\includegraphics{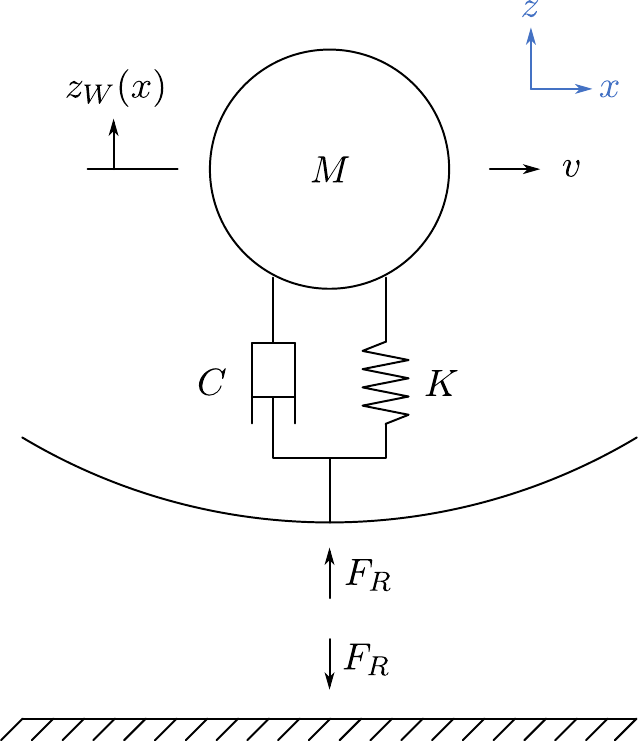}
	\caption{Dynamic model of the wheel/floor contact.}
	\label{fig:dynamic_model}
\end{figure}

The dynamic model, shown in \cref{fig:dynamic_model}, is based on a wheel modeled as a spring-mass-damper system resting on a rigid floor. The mass $M$ is the mass of the trolley plus added load, divided by the total number of wheels on the trolley. For a vehicle with a built-in suspension system, the stiffness used in the dynamic model is straightforward to calculate: it is simply the stiffness of the suspension spring. However, most trolleys do not have suspension systems: they contain a wheel turning about a rigid axle. In such a scenario, the stiffness used in the dynamic model is instead the stiffness of the wheel itself. The stiffness $K$ here is the equivalent elastic stiffness of the wheel, which is a function of the wheel's geometry and elastic properties. Polynomial relations were generated to estimate the wheel stiffness using a parametric finite element study \cite{edwardsPolynomialRelationsCylindrical2020}. A small amount of damping $C$ is added to be able to tune the model and avoid instabilities.

The wheel is represented by its receptance $G_W(f)$ and impulse response function $g_W(t)$.
\begin{equation} \label{eq:green_f}
G_W(f) = \frac{1}{-(2\pi f)^2 M + i(2\pi f) C + K}
\end{equation}
\begin{equation} \label{eq:green_t}
\tilde{g}_W(t) = \mathscr{F}^{-1}\left(G_W(f)\right)
\end{equation}
where $\mathscr{F}^{-1}$ represents the inverse Fourier transform.


The viscous damping ratio in the developed model was set to $\zeta = 0.4$. The viscous damping ratio and viscous damping factor are related via 
\begin{equation} \label{eq:zeta}
\zeta = \frac{C}{2\sqrt{KM}}
\end{equation}
This, in combination with the wheel stiffness, are included to ensure stability in the model (the model would otherwise have a tendency to go unstable when encountering moderate roughness asperities).

The equivalent spring foundation in the contact model calculates the deflection of the wheel and floor in the area of contact. This Hertzian evaluation does not account for the macro-level movement of the wheel (i.e. its center position $z_W$), thus it must be calculated separately and added in. To do so at a given $x$, a discretized form of the wheel Green's function is used \cite{pieringerFastTimedomainModel2008},
\begin{equation} \label{eq:g_w}
\begin{split}
& g_W(1) = \frac{1}{2}\Delta t\tilde{g}_W(0) \\
& g_W(n) = \Delta t\tilde{g}_W(n-1)\hspace{0.5cm} \text{for }n = 2,3,...,N-1 \\
& g_W(N) = \frac{1}{2}\Delta t \tilde{g}_W\left([N-1]\Delta t\right)
\end{split}
\end{equation}
where the discretization is defined by
\begin{equation} \label{eq:n}
t(n) = (n-1)\Delta t \hspace{0.5cm} \text{for }n = 1,2,...,N
\end{equation}
The change in vertical wheel position at step $n$ is calculated using the dynamic contact force and the discretized wheel Green's function.
\begin{equation} \label{eq:dz_w}
\Delta z_W(n) = g_W(1)(F_R(n) - Q)
\end{equation}
The vertical position of the wheel from the start of the rolling event up step $n$ is calculated as a convolution of the wheel Green's function and the dynamic contact force history (represented by the symbol $*$).
\begin{equation} \label{eq:z_wn}
z^{\textrm{old}}_W(0\xrightarrow\ n) = g_W*\left(F_R(0\xrightarrow\ n)-Q\right)
\end{equation}
The convolution is defined by
\begin{equation} \label{eq:conv_t}
\tilde{g}_W(t)*\left(F_R(0\xrightarrow\ t)-Q\right) = \int_{0}^{t}\left(F_R(\tau)-Q\right)\tilde{g}_W(t-\tau)\textrm{d}\tau
\end{equation}
Or the discretized version
\begin{equation} \label{eq:conv_n}
g_W(n)*\left(F_R(0\xrightarrow\ n)-Q\right) = \sum_{k=1}^{K}\left(F_R(k)-Q\right)g_W(n-k+1)
\end{equation}
Thus the vertical wheel position at step $n$ is
\begin{equation} \label{eq:z_w}
z_W(n) = \Delta z_W(n) + z^{\textrm{old}}_W(n)
\end{equation}
This results in an interdependence of \cref{eq:uprimer,eq:sigmar,eq:Fr,eq:z_w}. To alleviate this, the vertical wheel position at step $n$ is used to calculate the total interpenetration profile at step $n+1$. For the initial step $n=0$, a vertical wheel position of $z_W(0)=0$ is assumed.

Using this procedure, the wheel/floor interpenetration, contact force, and resulting wheel movement may be calculated for each position. At the end of the rolling event, the entire response history of the wheel and the floor is known.

\section{Discrete irregularities} \label{sec:discrete_irregularities}

\subsection{Wheel flats}
Wheel flats (or flat spots) are a possible artifact of indoor trolley wheels. A wheel which gets stuck for a period of time and slides across the floor may generate a flat spot due to friction. Alternatively, a large impact due to a drop or fall may cause a soft wheel to generate a flat spot as well. In both cases, the presence of the flat spot results in an impact occurring each time it completes a revolution, which has the potential to have a significantly higher sound level than that which is caused by the small scale roughness alone. 

\begin{figure}[!htbp]
	\centering
	\includegraphics{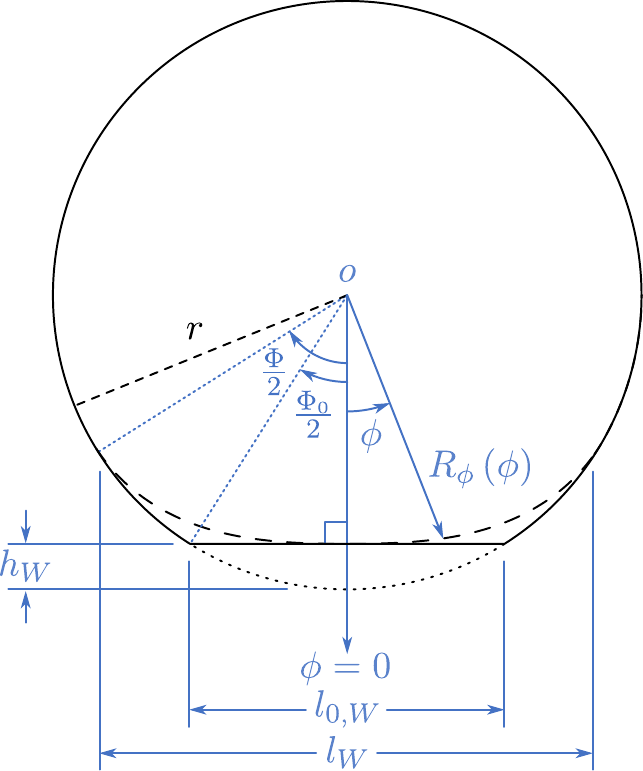}
	\caption{Geometry of an ideal and rounded wheel flat.}
	\label{fig:wheel_flat}
\end{figure}

Wheel flats are accounted for in the rolling noise model by modifying the wheel profile $\xi_W$ at each discretized step $n$. For a wheel of radius $r$, an ideal wheel flat may be modeled as a chord of the wheel circumference, as shown in \cref{fig:wheel_flat}. The flat depth $h_W$, ideal flat length $l_{0,W}$, and ideal center angle $\Phi_0$ are related by \cite{pieringerFastTimedomainModel2008}
\begin{equation} \label{eq:Phi0}
\frac{\Phi_0}{2} = \sin^{-1}\left(\frac{l_{0,W}}{2r}\right) = \cos^{-1}\left(1 - \frac{h_W}{r}\right)
\end{equation}
The profile of such a wheel may be described by
\begin{equation} \label{eq:Rphi0}
R_{0,\phi}(\phi) = \begin{cases}
r & \left|\phi\right| > \left|\frac{\Phi_0}{2}\right| \\
\frac{rcos\frac{\Phi_0}{2}}{\cos\phi} & \left|\phi\right| < \left|\frac{\Phi_0}{2}\right|  \\
\end{cases}
\end{equation}
where $\phi$ is evaluated over the interval $[-\pi,\pi]$.

Alternatively, some rolling noise models have had success using a rounded wheel flat profile, which replaces the hard corner transition between flat and curved with one which is more gentle \cite{pieringerFastTimedomainModel2008,baezaRailwayTrainTrackDynamics2006,wuHybridModelNoise2002}. This is intended to represent a wheel flat which has been in place for a period of time, and which has had its edges ``worn down'' from its ideal profile. Such a rounded profile may be represented by \cite{pieringerFastTimedomainModel2008}
\begin{equation} \label{eq:Phi}
\frac{\Phi}{2} = \sin^{-1}\left(\frac{l_W}{2r}\right) = \cos^{-1}\left(1 - \frac{h_W}{r}\right)
\end{equation}
\begin{equation} \label{eq:Rphi}
R_{\phi}(\phi) = \begin{cases}
r & \left|\phi\right| > \left|\frac{\Phi}{2}\right| \\
R - \frac{h_W}{2}\left[1 + \cos\left(\frac{2\pi\phi}{\Phi}\right)\right] & \left|\phi\right| < \left|\frac{\Phi}{2}\right|  \\
\end{cases}
\end{equation}

One may recall that the wheel radius $r_x$ is modified in the contact model in order to keep agreement with Hertzian contact mechanics and account for the non-linear deflection of the contact springs. Consequently, in order to map the flat spot to the wheel of reduced radius $r'_x$, the flat height is reduced here by the same amount.
\begin{equation} \label{eq:hwprime}
h'_W = h_W\frac{r'_x}{r_x}
\end{equation}

Similarly, the reduction of the wheel radius implicitly means that the wheel will make a greater number of revolutions across the floor for a given rolling event than what occurs in reality. This will result in an increase in the number of wheel flat impacts over a given rolling distance (in the case of a cylindrical wheel, the number of impacts will increase by a factor of two). This is corrected for by the following relation.
\begin{equation} \label{eq:dphi}
\Delta\phi = \frac{\Delta x}{r_x}
\end{equation}

At each discretized step $n$ in the model, the formulation of $\xi_W$ used in \cref{eq:uprimer} is updated as the wheel flat moves through the vicinity of the contact area. Care must thus be taken to evaluate the contact stress profile over a sufficiently large interval so as to capture the full length of the flat spot when it is in full contact with the floor (e.g. as it is shown in \cref{fig:wheel_flat}). In practice, this means that the interval [$-x',x'$] in the rolling noise model should be slightly longer than $2l_W$ when a wheel with flat spots is considered: much larger than the Hertzian contact length $2a$. Though the previous criteria still stands: the presence of any nonzero values on the edges of the stress matrix prior to integration of \cref{eq:Fr} indicates that the limits are too small.

\subsection{Floor joints}
\begin{figure}[!htbp]
	\centering
	\includegraphics{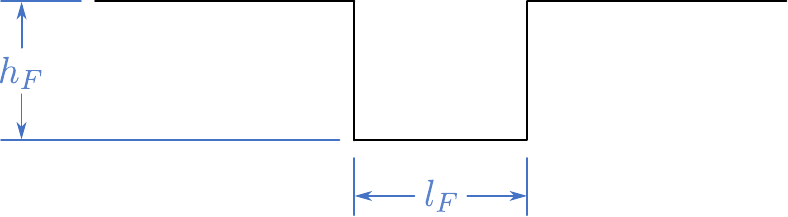}
	\caption{Geometry of a simple floor joint.}
	\label{fig:floor_joint}
\end{figure}

A similar approach is taken to account for the presence of floor joints in the rolling model. As shown in \cref{fig:floor_joint}, this is added directly into the rolling model by defining the floor profile in the global coordinate system ($x,z$) as the joint depth $-h_F$ within $\pm l_F/2$ of the joint center, and zero everywhere else. The joint centers are known based on the total rolling distance $L$ and the joint spacing $L_F$. Just as the wheel flat height was multiplied by $r'_x/r_x$ in order to correctly map the flat profile to the reduced wheel radius, the same procedure is applied here, but this time to the floor joint length $l_F$. The effect of the floor joint depth is accounted for in the model, but may be ignored in most cases. For all intents and purposes, the wheel will always contact the far side of the joint before ``bottoming out'' on the bottom of the joint. For such an event to occur, the wheel would need to be unreasonably small or the joint unreasonably long.

\section{Propagation model}\label{sec:propagation}
The transfer matrix method (TMM) is used in the field of acoustics to calculate the propagation of an acoustic wave through a multi-layered structure. This simplifies the problem greatly without compromising accuracy for scenarios where the propagation media are more or less homogeneous in two of the three dimensions. While examples of the use of the TMM in acoustics are plentiful for the case of airborne sound (e.g. \cite{foldsTransmissionReflectionUltrasonic1977,allardPropagationSoundPorous2009,brouardGeneralMethodModelling1995,atallaOverviewNumericalModeling2005}), examples of the use of the TMM for structure-borne sound are much more sparse \cite{geebelenModelEstimatingAcoustic2006}. Work was done by Rhazi and Atalla to develop formulae which use the TMM for mechanical excitation \cite{rhaziTransferMatrixModeling2010,rhaziSimpleMethodAccount2010}. The main limit here is the hypothesis of infinite lateral dimensions. The finite dimension is taken into account using the finite TMM (FTMM) \cite{villotPredictingAcousticalRadiation2001,rhaziTransferMatrixModeling2010}, which corrects for the radiation of the plate. However, the modal behavior of the floor is still not taken into account. It is these formulae which have been adapted for use in this rolling noise model.

The propagation model operates in the frequency domain to convert the previously calculated contact force and wheel response to a radiated sound power. This is done using \cite{rhaziTransferMatrixModeling2010}
\begin{equation} \label{eq:Pirad}
\Pi_{\textrm{rad}}=\frac{1}{8\pi^2}\int_{0}^{2\pi}\int_{0}^{\infty}\left|\frac{F_{\textrm{flex}}(f)}{F_{\textrm{ref}}}\right|^2\frac{Z_0\sigma_{\textrm{finite}}}{\left|Z_{\textrm{sTMM}}(f)+Z_{B,\infty}(f)\right|^2}k_r\textrm{d}k_r\textrm{d}\phi
\end{equation}
where $F_{\textrm{flex}}$  is the injected force, $F_{\textrm{ref}}$ is a reference force of 1 Newton, $Z_0$ is the acoustical impedance of air, $Z_{\textrm{sTMM}}$  is the impedance of the multilayered floor, $Z_{B,\infty} = k_0 Z_0 / \sqrt{k_0^2 - \left(k_x^2 + k_y^2\right)}$ is the radiation impedance seen from the excitation side \cite{fahySoundStructuralVibration2007}, and $k_0$ is the acoustic wavenumber. $\sigma_{\text{finite}}$ is the ``finite size'' radiation efficiency. The TMM is used to calculate the impedance of the multilayered floor $Z_{\textrm{sTMM}}$  for each wavenumber couple $\left(k_x,k_y\right)$ or $\left(k_r,\phi\right)$. In essence, the radiated power of an infinite plate is calculated, then spatially windowed to achieve the effective radiation of a plate with size equal to the actual floor.

The finite size radiation efficiency is given by
\begin{equation} \label{eq:sigma_finite}
\sigma_{\textrm{finite}} = \frac{\textrm{Re}(Z_R)}{Z_0S}
\end{equation}
where $S$ is the surface area of the multi-layer, and $Z_R$ is the radiation impedance considering the finite size. In reality, the finite area of the floor will play a role in it's radiation efficiency, thus the need for a correction term. This technique is sometimes referred to as \emph{spatial windowing}.

The injected flexural force $F_{\textrm{flex}}$ depends on the blocked force, which is taken to be the contact force $F_R(f)$. This takes into account the impedance of both the wheel and floor, using
\begin{equation} \label{eq:Fflex}
\left|F_{\textrm{flex}}(f)\right|^2=\left|\frac{Z_{\textrm{sTMM}}(f)}{Z_{\textrm{sTMM}}(f)+Z_W(f)}\right|^2\left|F_R(f)\right|^2
\end{equation}
Where $Z_W$ is the wheel impedance, estimated as the blocked force divided by the time derivative of the vertical wheel position. 
This is more robust than calculating the wheel impedance from the receptance model, as it includes the effect of both the relative roughness and the contact stiffness.


As with the wheel impedance, the floor impedance may theoretically be calculated by dividing the injected force by the floor velocity. However, as the injected force in calculated from the contact force found in the contact model, it may appear that it cannot be known ahead of time. To alleviate this problem, the floor impedance is pre-calculated using an external point force of 1~N. This essentially yields a unit floor impedance. When used in \cref{eq:Fflex} in conjugation with the blocked force, this will yield the correct injected flexural force.

The presence of a floor covering is accounted for in both the contact model (via the different roughness profile) and the propagation model (via the inclusion of the floor covering as an added layer in the TMM).

The normalized sound pressure level $L_n$ is determined from the radiated sound power using the process determined in ISO 10140-3 \cite{ISO10140320102010}
\begin{equation} \label{eq:Ln_model}
L_n = 10\log\frac{\Pi_{\textrm{rad}}}{\Pi_{\textrm{ref}}} + 10\log\frac{4}{A_0} = 10\log_{10}\frac{I_{\textrm{rad}}}{\Pi_{\textrm{ref}}} + 10\log\frac{4S}{A_0}
\end{equation}
where $S$ is the surface area of the floor in square meters, $A_0 = 10~\textrm{m}^2$ is the reference surface area, and $\Pi_\textrm{ref} = 10^{-12}$~W is the reference sound power. \cref{eq:Pirad} is calculated using a sphere of unit surface area, making it equal to the sound intensity. It may thus also be used here to directly calculate $L_n$ by including the true surface area in \cref{eq:Ln_model}. The normalized sound pressure level was chosen as the quantity of interest because it allows for direct comparison with impact noise, being a part of an already existing standard.

\section{Roughness measurement} \label{sec:roughness}
In order to provide an accurate estimation of the relative roughness between the wheel and the floor as input to the rolling noise model, real roughness profiles are needed. This presents a particularly difficult problem for indoor rolling noise. Roughness profiles measured for train wheel/rail and vehicle tire/road contact are typically done with a spatial resolution of 0.5--1~mm (e.g. as in \cite{pieringerInvestigationDynamicContact2011}). This is acceptable for the large contact dimensions at play in these scenarios. However, due to the smaller wheel sizes and static loads with indoor trolleys, where the length of the contact area in the longitudinal direction is often on the order of 1--2~mm, a much higher spatial resolution is necessary.

For this model, roughness samples from one cylindrical wheel and three floors were measured. This was done using a Nikon LC15DX 3D scanner with an average spatial resolution of $22~\mu$m. The floors measured were polished concrete, a smooth PVC floor covering, and a rough PVC floor covering. The two PVC floors were identical in construction, the only difference being their surface roughness. A 3.5~cm wide section was measured along the full length of each floor sample: 20~cm for the concrete, and 60~cm for the PVC. For the wheel measurement, the entire rolling surface was captured, giving a width equal to the wheel width (3.5~cm), and length equal to the wheel's circumference ($10\pi$~cm).

The roughness profiles were incorporated into the model by starting with an array containing the measured profile of a given floor or wheel: low-pass filtered with a cutoff wavelength of 20~mm (which corresponds to a frequency of 50~Hz for a trolley speed of 1~m/s). The profiles were duplicated in the rolling direction using a mirroring technique until the roughness profile length was equal to or greater than the desired rolling length $L$ specified in the model, then truncated to be exactly equal to $L$.

Each time the profile was duplicated, the added section was mirrored, such that the first value in the array becomes the last and the last value in the array becomes the first, before adding it to the end of the existing profile. This technique eliminates the need to window the profile sections in order to avoid a discontinuity in profile height from section to section, which would otherwise result in areas of smaller roughness magnitude periodically throughout the total profile. The same procedure (duplicating using the mirroring technique, then truncating to the desired size) was used to extend the roughness profiles to the desired width in the transverse direction as well.

This process means that a single roughness profile exists in the model for a given wheel or floor type. The relative roughness profile between the wheel and floor will, however, change depending on the ratio of wheel circumference to floor section length. No explicit variations are incorporated into the model's handling of the roughness profile.

\section{Experimental results} \label{sec:experimental}
The model was compared to experimental results gathered from a rolling noise test performed at Level Acoustics \& Vibration in Eindhoven, the Netherlands. The sound generated by a rolling test trolley was captured in a two-story transmission room: with the trolley being rolled in the top room, and the sound measured in both the top (emission) and bottom (reception) rooms. The measurement procedure outlined in ISO 10140-3:2010 \cite{ISO10140320102010} for laboratory testing of a tapping machine was followed in terms of microphone placement and calculation of the normalized sound pressure level $L_n$. The 3~m $\times$ 3~m opening in the floor between the two transmission rooms was filled with a 100~mm thick concrete slab. This slab was decoupled from the surrounding floor to eliminate sound transmitted to the reception room via flanking transfer paths. The emission and reception rooms each had a volume of 80 $\textrm{m}^3$.

\begin{figure}[!htbp]
	\centering
	\includegraphics[width=0.25\textwidth]{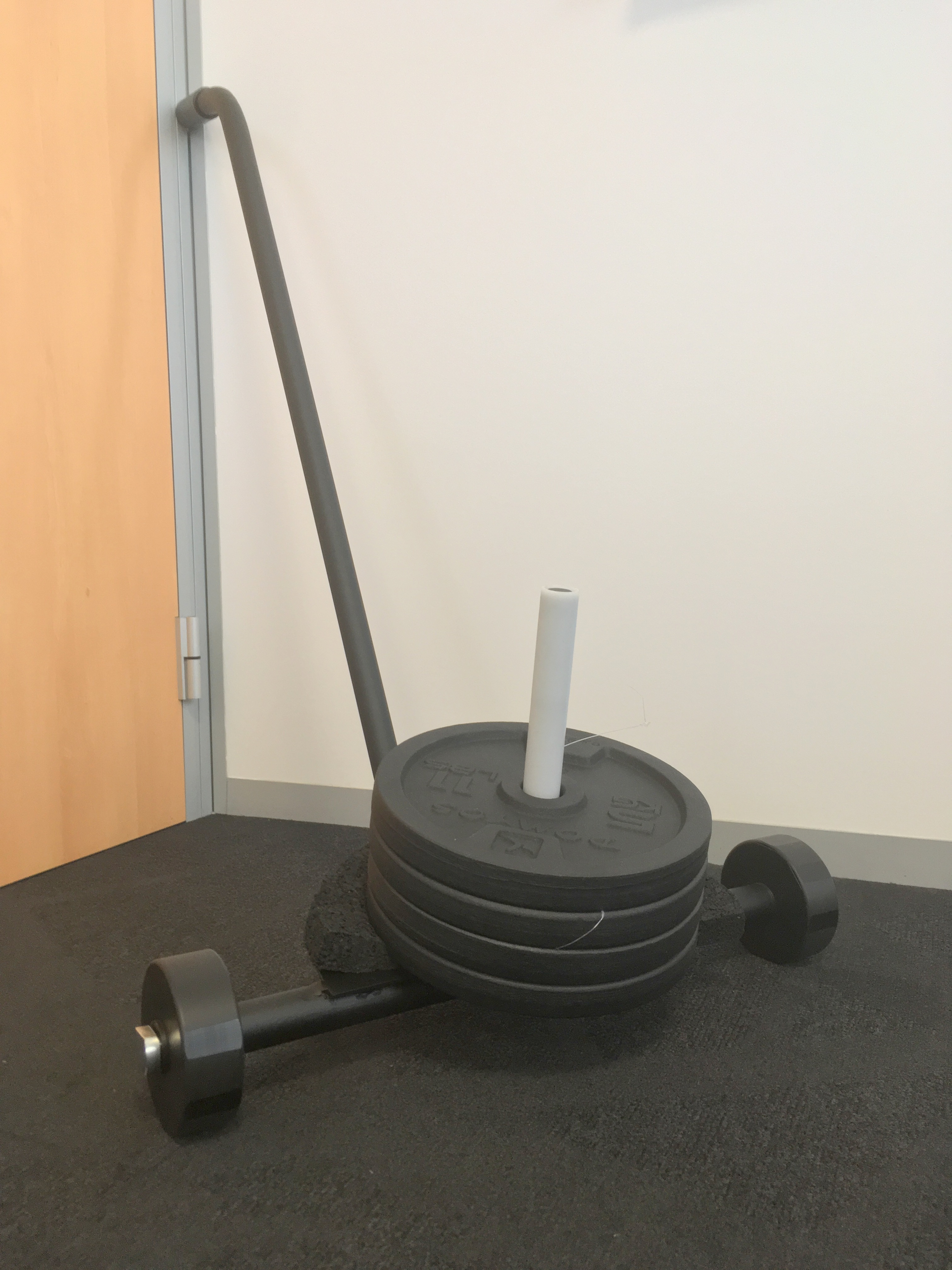}
	\caption{Two-wheeled test trolley used in the rolling noise experiments. Shown here with flat wheels and a 20~kg added load.}
	\label{fig:test_trolley}
\end{figure}

The test trolley, shown in \cref{fig:test_trolley} used had a simple two-wheel design, as it allows the rolling noise itself to be accurately represented while eliminating all other sources of noise which are beyond the scope of this model (e.g. rattling noise). Two types of wheels were used on the trolley with each of the three floors: a smooth cylindrical plastic wheel and an identical wheel which had six non-periodic flat spots added around it's perimeter. The flat spots each had a depth of 0.5~mm (compared to a wheel radius of 50~mm).

For each of these tests, the trolley was rolled back and forth repeatedly in a linear trajectory for a period of 30 seconds, and the sound averaged over the entire time span for each run. A total of five runs were performed and averaged together for each unique configuration in order to elimination potential variations between runs.

The varying parameters of the test were the floor type, the wheel type, the trolley speed, and the added load on the trolley. These are summarized in \cref{tab:test_params}. Three different weight configurations were used: 10, 20, and 30~kg. These added loads were all in addition to the 7~kg mass of the trolley itself.
\begin{table}[!htbp]
	\centering
	\caption{Test parameters used in the rolling noise test at Level Acoustics \& Vibration} \label{tab:test_params}
	\begin{tabular}{l l l l}
		\hline
	Floor	&	Wheel	&	Load	&	Speed	\\
		\hline 
	Concrete	&	Smooth	&	10 kg	&	0.5 m/s	\\
	Concrete + rough PVC	&	Flat	&	20 kg	&	0.9 m/s	\\
	Concrete + smooth PVC	&		&	30 kg	&	1.6 m/s	\\
		\hline
	\end{tabular}
\end{table}

The experimental tests showed very low variation between the five repeated runs for a given configuration. Average standard deviations of the normalized sound pressure level $L_n$ were generally less than or equal to 1~dB, with only one configuration (rough PVC floor, smooth wheel, 0.9~m/s trolley speed, and 10~kg added load) having a slightly higher standard deviation of 1.3~dB.

In the following section, a number of comparisons will be made between model and experimental results. There are a large number of variables which may change from run to run, so for the purpose of brevity, several default values will be first declared here. Unless otherwise stated, the parameters shown in \cref{tab:default_params} are used for the results shown in \cref{sec:model_validation}. If a particular run deviates from these default parameters (e.g. if a 20~kg added load is used instead of a 10~kg added load), it will be specified as such. The wheel damping ratio $\zeta$ and the viscous damping factor $C$ are related by \cref{eq:zeta}. The spatial resolution d$x$ in particular was chosen to yield a maximum frequency above 5~kHz (thus a sampling frequency of 10~kHz) for the lowest trolley speed considered (0.5 m/s).
\begin{table}[!htbp]
	\centering
	\caption{Default parameters used for the results shown in \cref{sec:experimental}, unless otherwise specified.}
	\label{tab:default_params}
	\begin{tabular}{l c c S}
		\hline
		Parameter	&	Variable	&	Unit	&	\multicolumn{1}{c}{Value}	\\
		\hline
		Spatial resolution  &   d$x$    &   ($\mu$m)  &  55  \\
		Wheel radius	&	$r_x$	&	(mm)	&	50	\\
		Wheel transverse radius	&	$r_y$	&	(mm)	&	$\infty$	\\
		Wheel width	&	$w$	&	(mm)	&	35	\\
		Wheel damping ratio	&	$\zeta$	&	(--)	&	0.4	\\
		Wheel flat depth (if applicable)	&	$h_W$	&	(mm)	&	0.5	\\
		Floor joint length (if applicable)	&	$l_F$	&	(mm)	&	10	\\
		Floor joint depth (if applicable)	&	$h_F$	&	(mm)	&	2.5	\\
		Floor joint spacing (if applicable) &   $L_F$   &   (m)     &   0.33 \\
		Trolley speed	&	$v$	&	(m/s)	&	0.9	\\
		Rolling distance	&	$L$	&	(m)	&	1	\\
		Added load	&	$M_l$	&	(kg)	&	10	\\
		Trolley mass	&	$M_t$	&	(kg)	&	7	\\
		\hline
	\end{tabular}
\end{table}

The plastic test wheel and three flooring materials (polished concrete, rough PVC, and smooth PVC) have material properties given in \cref{tab:material_props}. They were obtained from the respective material manufacturers. These material characteristics are known to provide good results (measured vs calculated) when evaluating impact noise. The dynamic damping loss factor is used for the floors in the model. The mean value across the entire frequency range is shown here for brevity.
\begin{table}[!htbp]
	\centering
	\caption{Material properties for the wheel and floors used for the results shown in \cref{sec:experimental}.} \label{tab:material_props}
	\begin{tabular}{l c c c}
		\hline
		Parameter	&	Plastic wheel    &   Polished concrete    &   PVC floor   \\
		&   &   floor   & covering  \\
		\hline
		Young's modulus $E$ (GPa) &   0.2 &   33  &   0.75    \\
		Poisson's ratio $\nu$ (--) &   0.3 &   0.3 &   0.3 \\
		Mass density $\rho$ (kg/$\textrm{m}^3$) &   1428    &   2150    &   1556    \\
		Mean damping loss factor $\bar{\eta}$ (--)  &   --  &   0.32    &   0.02    \\
		\hline
	\end{tabular}
\end{table}

\section{Model validation} \label{sec:model_validation}
The following sections show comparisons between model and experimental results through various influencing parameter changes: the floor type, the presence (or not) of wheel flat spots, the load on the trolley, and the speed of the trolley. Intermediary model results are also presented for the contact force, contact force spectrum, wheel impedance, and floor admittance. Concerning the wheel flat spots: the model makes predictions for two types of flat spot profiles: ideal and rounded. However, only one type of flat wheel was used in experimental testing: being neither perfectly ideal nor perfectly rounded in profile. Model results for both the ideal and rounded flat wheel are compared to experimental results for the one type of real flat wheel in order to identify which model profile yields a better approximation.

\subsection{Normalized sound pressure level} \label{sec:modelLn}
\subsubsection{Influence of the floor and wheel type} \label{sec:compFloorWheel}
\begin{figure}[ht!]
	\centering
	\subfloat[][]{\includegraphics{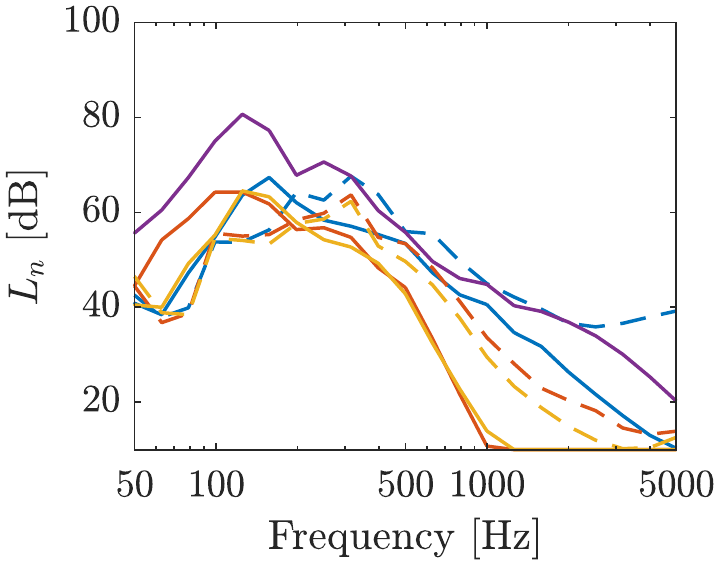}}\\
	\subfloat[][]{\includegraphics{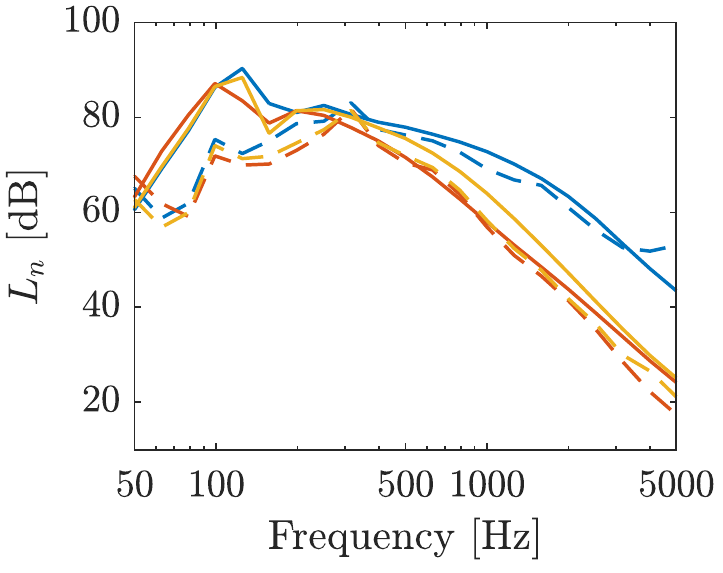}}
	\subfloat[][]{\includegraphics{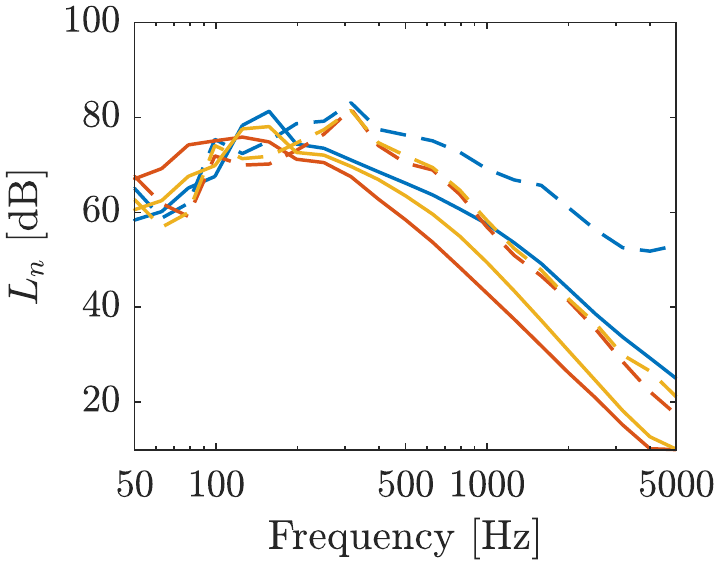}}
	\caption{Normalized sound pressure level in the reception room for varying floor covering: model versus experiment. (a) Smooth wheel, (b) ideal flat wheel, (c) rounded flat wheel.
	\mbox{\protect\includegraphics{line1} model,} smooth concrete floor, 
	\mbox{\protect\includegraphics{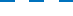} experiment,} smooth concrete floor, 
	\mbox{\protect\includegraphics{line2} model,} rough PVC floor, 
	\mbox{\protect\includegraphics{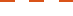} experiment,} rough PVC floor, 
	\mbox{\protect\includegraphics{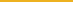} model,} smooth PVC floor, 
	\mbox{\protect\includegraphics{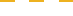} experiment,} smooth PVC floor, 
	\mbox{\protect\includegraphics{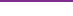} model,} jointed concrete floor.}
	\label{fig:Ln_compFloor}
\end{figure}

\cref{fig:Ln_compFloor} shows the normalized sound pressure level calculated by the rolling noise model versus experimental results for the smooth wheel, ideal flat wheel, and rounded flat wheel on three floors: smooth concrete, concrete + rough PVC covering, and concrete + smooth PVC covering. In addition, model results for the smooth wheel rolling on a theoretical jointed concrete floor are also shown.

The concrete floor generally yields higher levels than the rough and smooth PVC floors, whose curves are similar to one another in shape and magnitude. Model results agree well above 300~Hz for the ideal flat wheel. The smooth wheel and rounded flat wheel tend to underpredict across the whole frequency range, however their general shapes are still congruent with those of their respective experimental curves. The average background noise level in the high frequency range for the measurements corresponded to 10~dB, so any model results below this were corrected to this level. This can be seen particularly in the results for the smooth wheel above 1~000~Hz.

The case of the jointed floor was not able to be measured experimentally. The model results are shown in \cref{fig:Ln_compFloor} to demonstrate the model's ability to handle floor joints. The sound level predicted by the model for the smooth wheel rolling on a jointed concrete floor is similar to that of the rounded flat wheel on a smooth concrete floor in the range of 100~Hz. Reduced levels (8 dB lower on average) are predicted in the range of 200--5~000~Hz. Considering, in addition, that the model results for the smooth wheel on jointed concrete are roughly the same in the range of 300--2~000~Hz as the experimental results for the smooth wheel on a smooth concrete floor, it is likely that the model is underpredicting for the case of floor joints.

\begin{figure}[ht!]
	\centering
	\subfloat[][]{\includegraphics[scale=0.99]{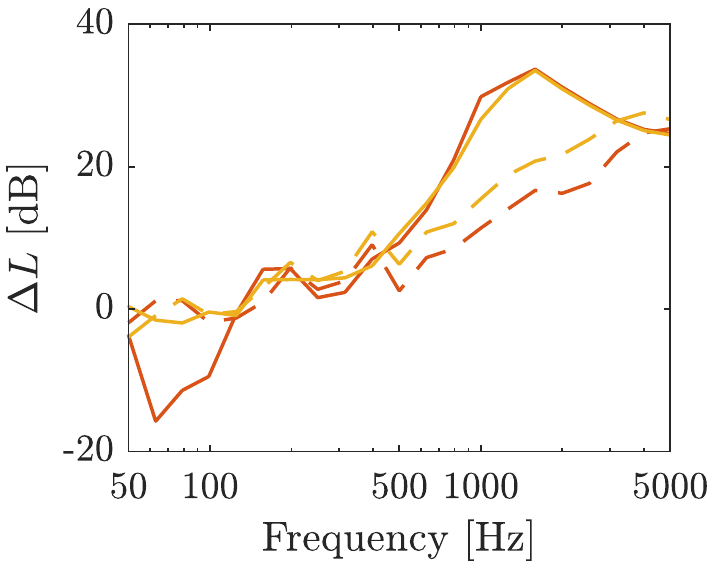}}\\
	\subfloat[][]{\includegraphics[scale=0.99]{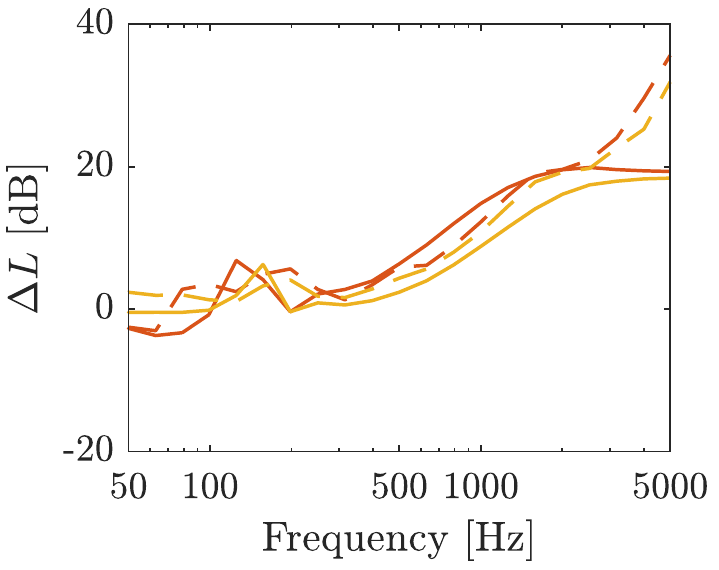}}
	\subfloat[][]{\includegraphics[scale=0.99]{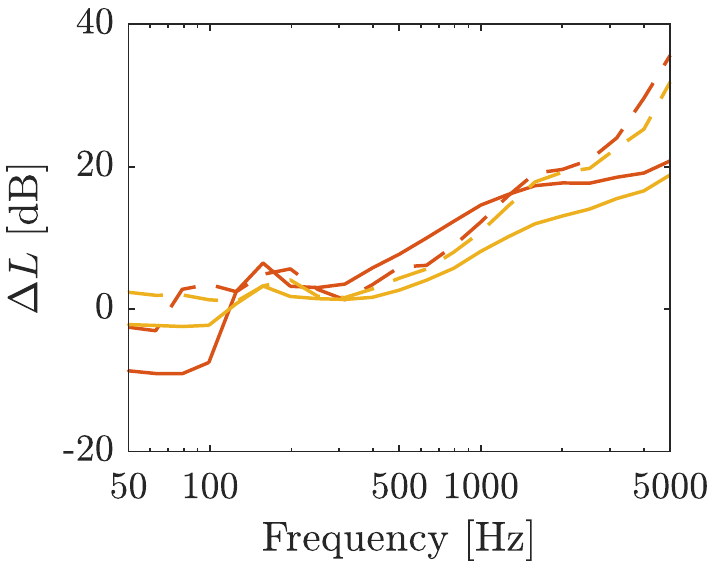}}
	\caption{Benefit from applying the rough and smooth PVC floor coverings: model versus experiment. (a) Smooth wheel, (b) ideal flat wheel, (c) rounded flat wheel. 
	\mbox{\protect\includegraphics{line2} model,} rough PVC floor, 
	\mbox{\protect\includegraphics{line6} experiment,} rough PVC floor, 
	\mbox{\protect\includegraphics{line3} model,} smooth PVC floor, 
	\mbox{\protect\includegraphics{line7} experiment,} smooth PVC floor.}
	\label{fig:deltaL_compFloor}
\end{figure}

\cref{fig:deltaL_compFloor} shows the benefit of adding a PVC floor covering onto the smooth concrete floor, as computed with both model and experimental data via $\Delta L$. There is quite a bit less difference between the three wheel types in the $\Delta L$ plots than in the $L_n$ plots, indicating that the floors provide more or less a similar benefit regardless of wheel type. The model estimates the relative sound level $\Delta L$ more accurately than the absolute sound level $L_n$. It is worth noting that while the model underpredicts in \cref{fig:Ln_compFloor} with the rounded flat wheel, it still provides a reasonably accurate prediction of $\Delta L$ in \cref{fig:deltaL_compFloor}. The combination of a smooth wheel (excitation via only roughness asperities) and a soft floor (PVC) results in little excitation in the model. Thus, other model parameters beyond the roughness, wheel, and floor profiles have a greater relative influence on the results. While the dynamic loss factor (used in the TMM to calculate the floor admittance) in 1/3 octave bands is well known for concrete, the dynamic loss factor of the PVC floor coverings was only measured at two frequencies: 105 and 698~Hz. Values for the remainder of the frequency range were extrapolated from these two points. Any deviation between the estimated and true dynamic loss factor will thus manifest itself as a greater deviation between model and experimental results when the excitation is low. The model is more robust at predicting configurations with higher excitation (i.e. harder materials and/or larger asperities).

The smooth PVC and rough PVC curves differ slightly from one another (0--2 dB), for both model and experimental results. This difference is greater in the model results for the ideal and rounded flat wheels than the smooth wheel. The model is particularly sensitive to the relative roughness profile in the presence of discrete irregularities. The model predicts that larger roughness asperities will result in a lower impact force when the flat spot falls to impact with the floor below: possibly due to there being fewer discretized points in contact with one another at the moment of impact. Nevertheless, all curves show a greater $\Delta L$ over the concrete. This indicates that the attenuation observed in \cref{fig:deltaL_compFloor} is primarily due to the elasticity of the PVC layer, rather than the difference in surface roughness.

As an aid in interpreting the effect of discrete irregularities on the radiated sound power, we may visualize a one-dimensional approximation of the relative wheel/floor interpenetration profiles. This is done by modifying  \cref{eq:uprimer} to be
\begin{equation} \label{eq:uprimer_eq}
u_{R,\textrm{eq}}(x) = \varsigma(x) - \xi_W(x) + \xi_F(x)
\end{equation}
which is now a function of only $x$. Here $\varsigma(x)$ is taken to be the center line of the three dimensional relative roughness profile from \cref{eq:uR}. $\xi_W(x)$ and $\xi_F(x)$ are found by forming a new array with the center points of the wheel and floor profiles $\xi_W(x',y')$ and $\xi_F(x')$ for each $x$. This one-dimensional approximation cannot fully represent the true interpenetration profile, as it cannot fully account for the relationship between the position of the wheel flat and the curvature of the wheel at a given $x$. Nevertheless, it can be used to help visualize and understand the effect of the discrete irregularities (and to a lesser extent, the small scale roughness) in both the spatial and frequency domains.

\begin{figure}[!htbp]
	\centering
	\subfloat[][]{\includegraphics{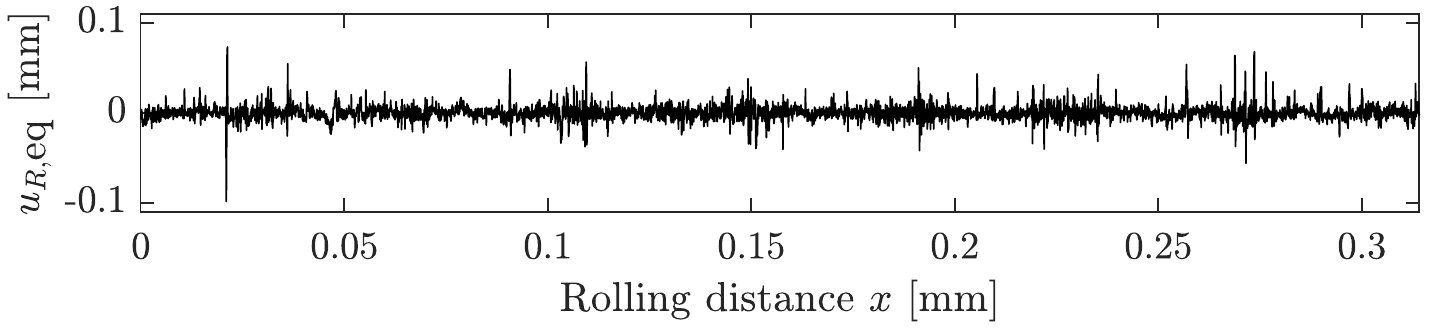}}\\
	\subfloat[][]{\includegraphics{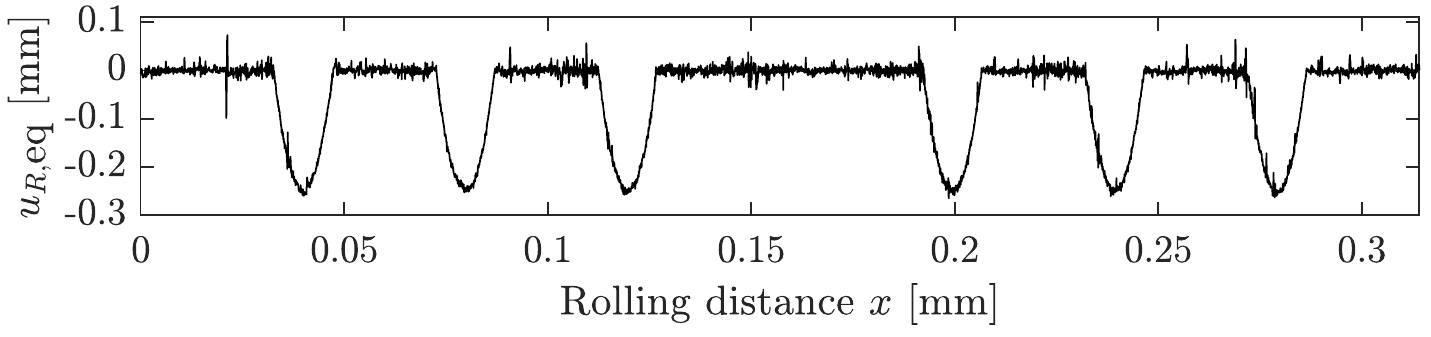}}\\
	\subfloat[][]{\includegraphics{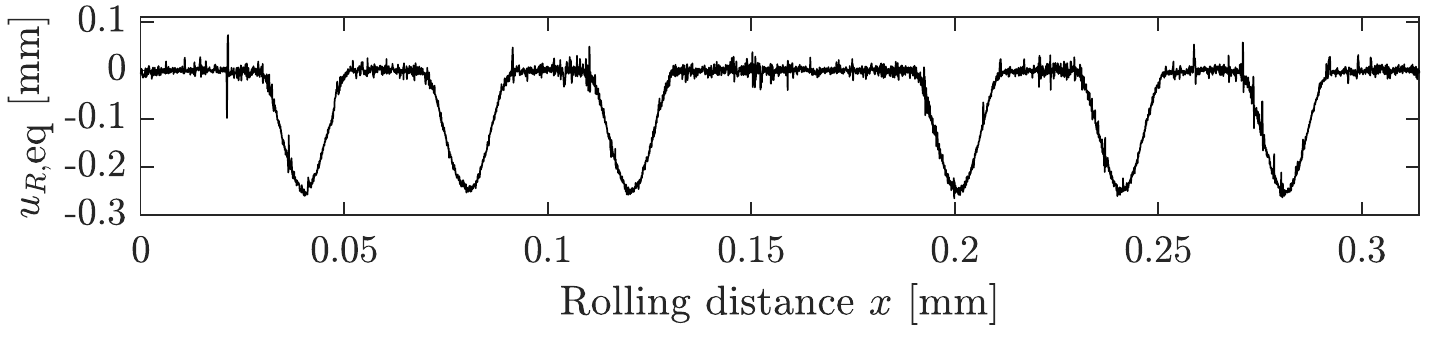}}\\
	\hspace{0.4cm}\subfloat[][]{\includegraphics{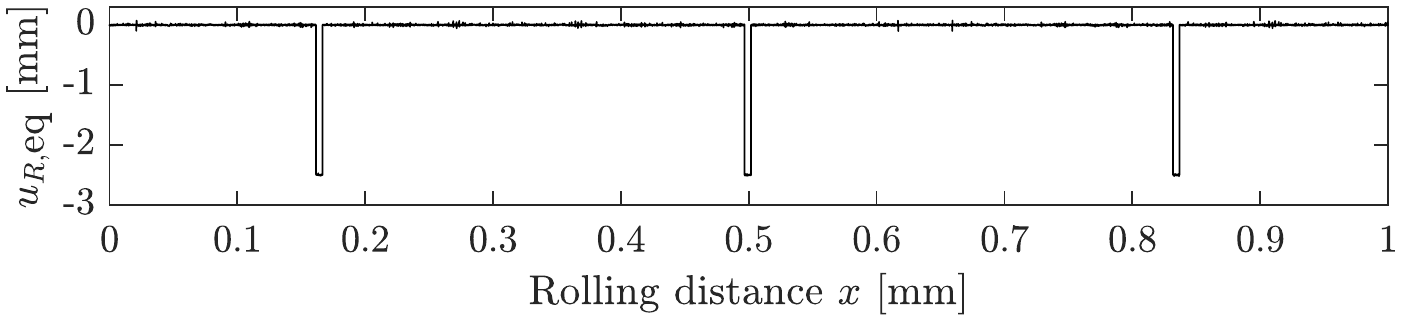}}
	\caption{One-dimensional approximation of the wheel/floor interpenetration profiles: (a) Smooth wheel, smooth concrete floor, (b) ideal flat wheel, smooth concrete floor, (c) rounded flat wheel, smooth concrete floor, (d) smooth wheel, jointed concrete floor.}
	\label{fig:rough_uReq_pos}
\end{figure}

\cref{fig:rough_uReq_pos} shows four different approximated 1D interpenetration profiles: a smooth wheel rolling on smooth concrete, an ideal flat wheel on smooth concrete, a rounded flat wheel on smooth concrete, and a smooth wheel on jointed concrete. The three smooth concrete profiles are shown for a rolling distance of one wheel circumference. The jointed concrete profile is shown for a rolling distance of 1~m.

\begin{figure}[!htbp]
	\centering
	\includegraphics{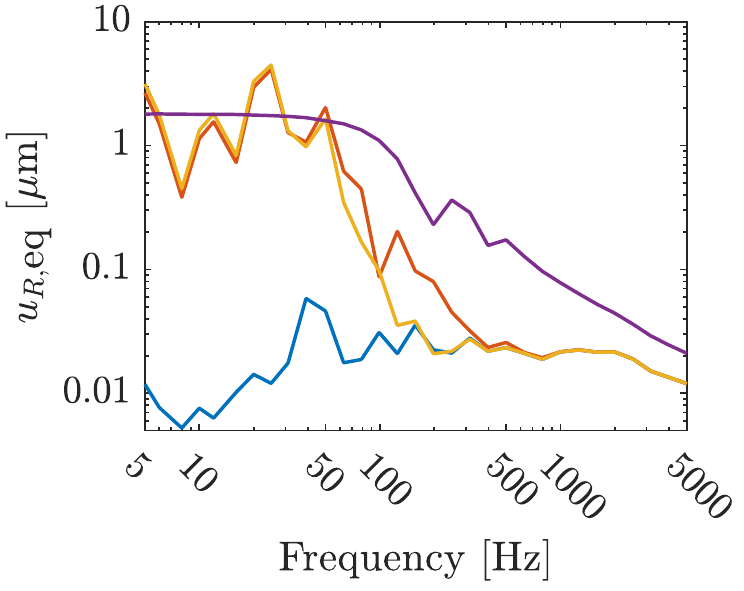}
	\caption{Frequency spectra of the one-dimensional approximation of the wheel/floor interpenetration profiles. 
	\mbox{\protect\includegraphics{line1} Smooth} wheel, smooth concrete floor, 
	\mbox{\protect\includegraphics{line2} ideal} flat wheel, smooth concrete floor, 
	\mbox{\protect\includegraphics{line3} rounded} flat wheel, smooth concrete floor, 
	\mbox{\protect\includegraphics{line4} smooth} wheel, jointed concrete floor.}
	\label{fig:rough_uReq_spect}
\end{figure}

\cref{fig:rough_uReq_spect} shows the frequency spectra of the approximated 1D profile in 1/3 octave bands for a trolley speed of 0.9~m/s. The primary fundamental frequencies of the wheel flat impacts are visible in the 10~Hz and 25~Hz 1/3 octave bands (as well as a harmonic at 50 Hz). It is also worth noting here that the overall magnitude (but not the shape) of the jointed concrete spectrum is overestimated here: in reality the wheel will not contact the bottom of the 2.5~mm deep floor joint. However, in the frequency range of interest for rolling noise (50--1~000~Hz), there do not appear to be any artifacts which would directly explain the trend in the model $L_n$ curves to exhibit a peak around 100~Hz.

\subsubsection{Influence of the added load} \label{sec:compMass}
\begin{figure}[ht!]
	\centering
	\subfloat[][]{\includegraphics{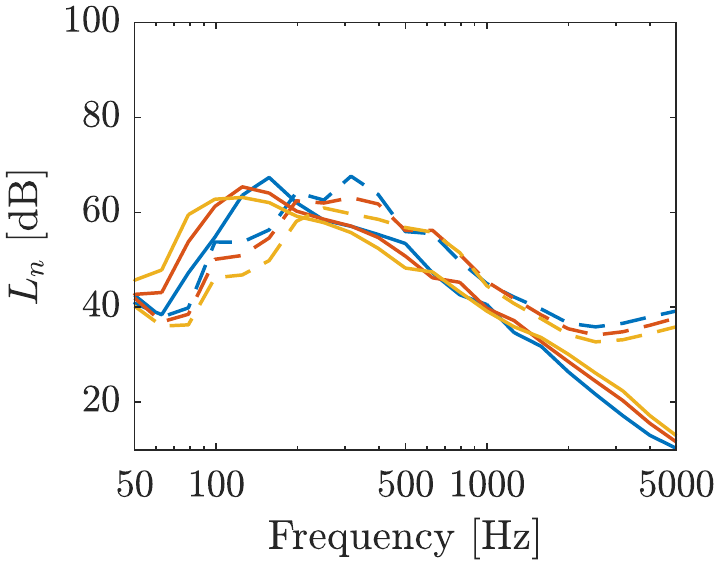}}\\
	\subfloat[][]{\includegraphics{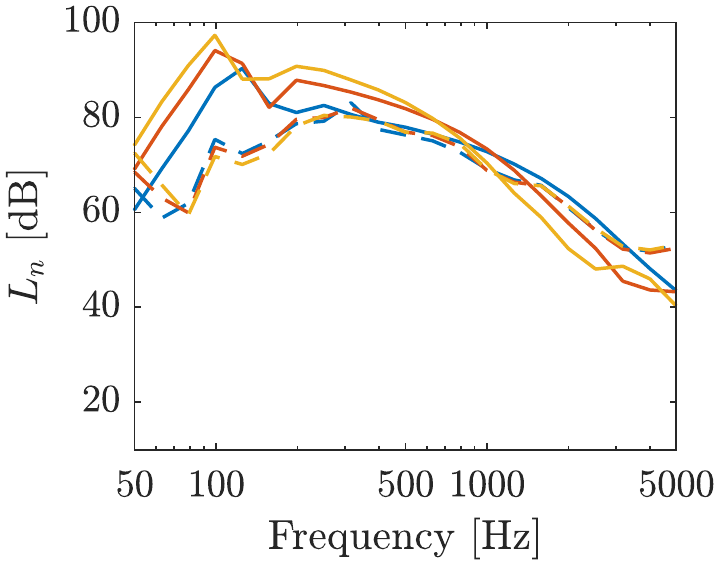}}
	\subfloat[][]{\includegraphics{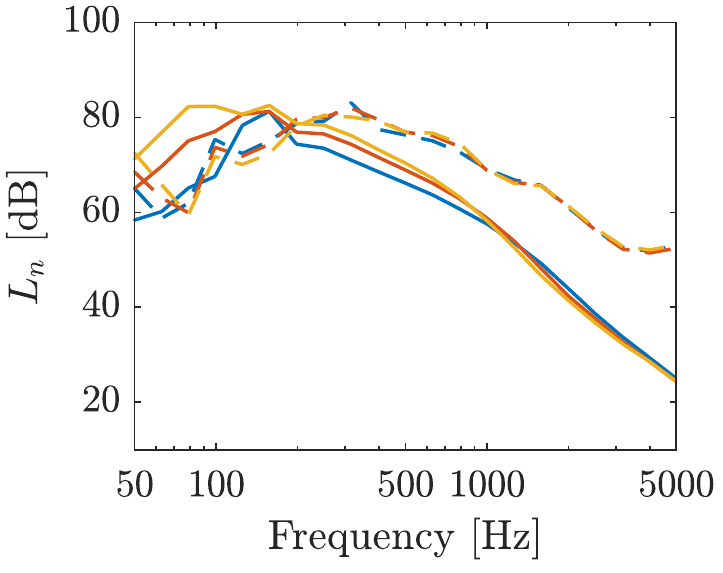}}
	\caption{Normalized sound pressure level in the reception room for varying added load: model versus experiment. For all curves, the floor was smooth concrete. (a) Smooth wheel, (b) ideal flat wheel, (c) rounded flat wheel.
	\mbox{\protect\includegraphics{line1} Model,} 10~kg, 
	\mbox{\protect\includegraphics{line5} experiment,} 10~kg, 
	\mbox{\protect\includegraphics{line2} model,} 20~kg, 
	\mbox{\protect\includegraphics{line6} experiment,} 20~kg, 
	\mbox{\protect\includegraphics{line3} model,} 30~kg, 
	\mbox{\protect\includegraphics{line7} experiment,} 30~kg.}
	\label{fig:Ln_compMass}
\end{figure}

\cref{fig:Ln_compMass} shows the influence of changing the added load on the trolley in both the model and the experimental results. For all curves, the floor was smooth concrete. The model correctly predicts only a slight change in sound level with changing wheel load. However in the low frequency range, where an inverse relationship is expected (i.e. a slight decrease in sound level with increasing added load), the effect is not seen in the model results. Instead, a positive relationship is predicted. As the load increases, new roughness asperities come into contact with one another which were out of contact at lower loads, resulting in a low frequency shift. This phenomenon is present in the experimental results, but might not be fully captured by the model. It may be due to the use of an independent bed of springs rather than elastic half spaces, which account for the elastic deformation of the contact surfaces due to roughness asperities \cite{boussinesqApplicationPotentielsEtude1885}.

\begin{figure}[ht!]
	\centering
	\subfloat[][]{\includegraphics{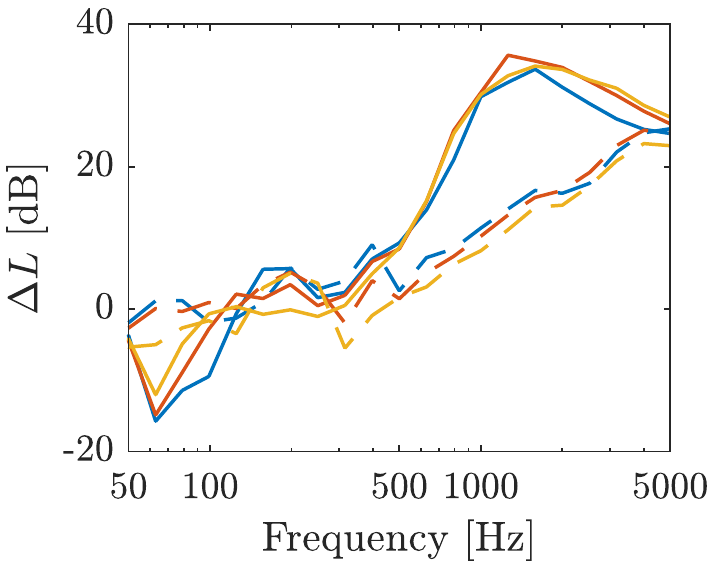}}\\
	\subfloat[][]{\includegraphics{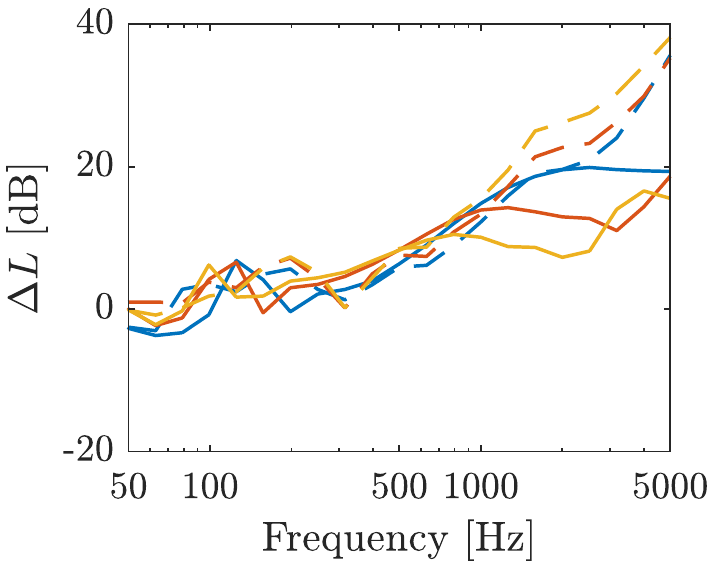}}
	\subfloat[][]{\includegraphics{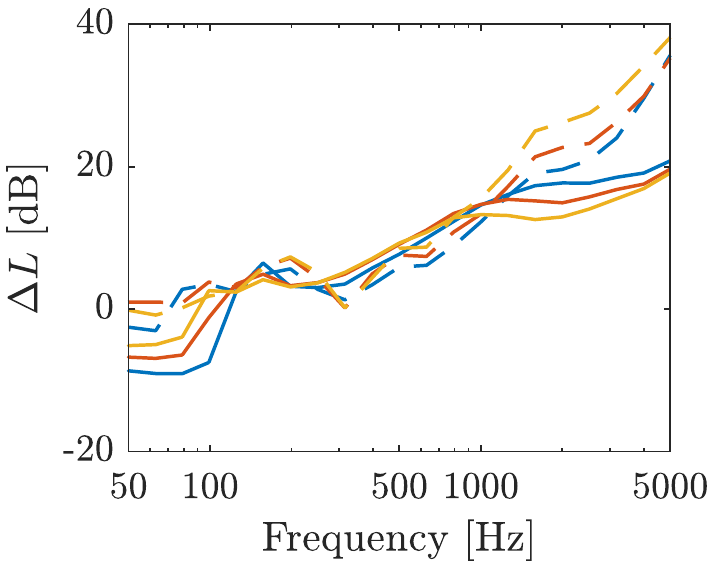}}
	\caption{Benefit from applying the rough PVC floor covering for varying added load: model versus experiment. (a) Smooth wheel, (b) ideal flat wheel, (c) rounded flat wheel. 	\mbox{\protect\includegraphics{line1} Model,} 10~kg, 
	\mbox{\protect\includegraphics{line5} experiment,} 10~kg, 
	\mbox{\protect\includegraphics{line2} model,} 20~kg, 
	\mbox{\protect\includegraphics{line6} experiment,} 20~kg, 
	\mbox{\protect\includegraphics{line3} model,} 30~kg, 
	\mbox{\protect\includegraphics{line7} experiment,} 30~kg.}
	\label{fig:deltaL_compMass}
\end{figure}

\cref{fig:deltaL_compMass} shows how the benefit of adding a rough PVC floor covering changes with trolley added load for both model and experimental results. The results are similar to those shown in \cref{fig:deltaL_compFloor}. Here we see that the model predicts that the floor covering will perform slightly worse as the load increases for the case of the flat wheel. However this occurs most strongly in the high frequency region, where there is little acoustic energy. The mid and low frequency regions tend to agree better with experimental results (particularly for the flat wheels).

The dynamic system used in this model implicitly assumes that the mass of the trolley is equally distributed over all wheels and that there is no rotation of the trolley body. The validity of these assumptions depends primarily on the construction of the trolley. A trolley whose center of gravity differs from its center of mass will have wheels which do not bare an equal proportion of the weight. This unequal distribution may result in rotation of the trolley body during rolling, which cannot be simulated using the spring-mass-damper system used in this model. The effect of the added load on the trolley, however, has been shown to have a minimal effect on the radiated sound level (a difference of 10 kg in the added load only results in a 0--2~dB difference in the overall sound level). Therefore, any effect of unequal mass distribution on the trolley wheels is assumed here to be negligible.

\subsubsection{Influence of trolley speed} \label{sec:influence_speed}
\begin{figure}[ht!]
	\centering
	\subfloat[][]{\includegraphics{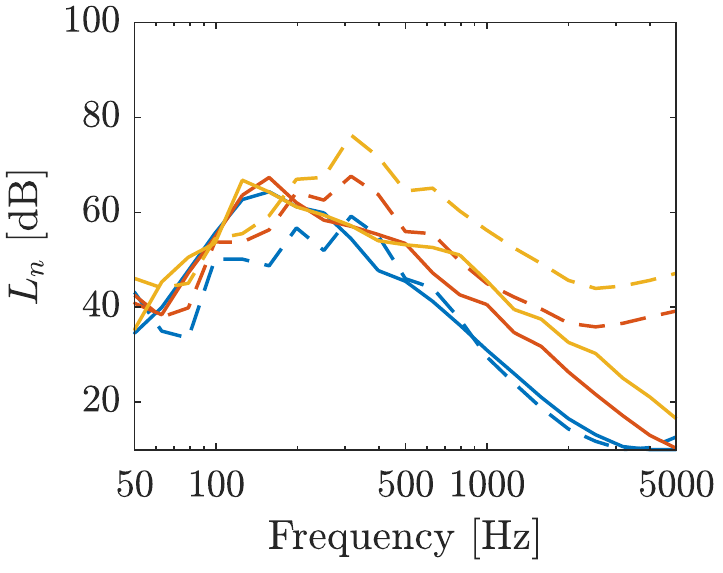}}\\
	\subfloat[][]{\includegraphics{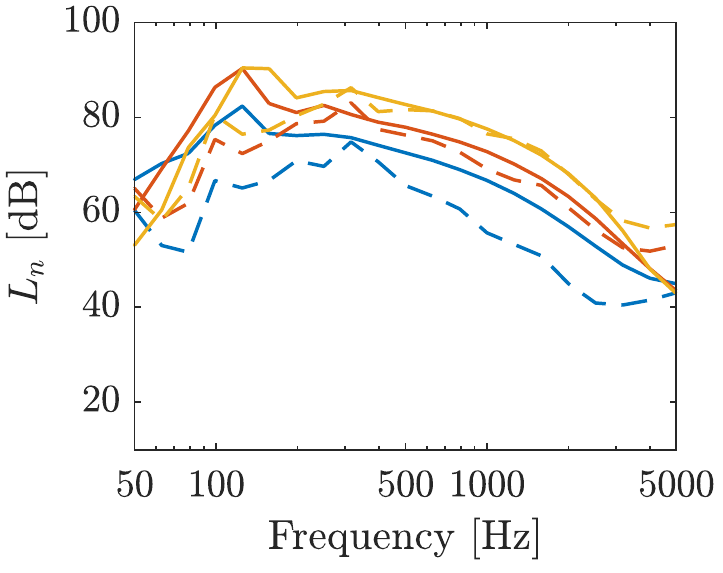}}
	\subfloat[][]{\includegraphics{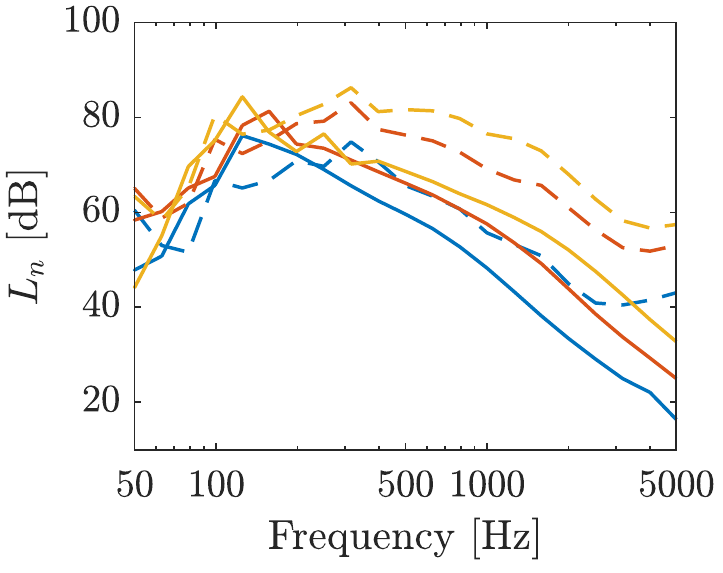}}
	\caption{Normalized sound pressure level in the reception room for varying trolley speed: model versus experiment. For all curves, the floor was smooth concrete. (a) Smooth wheel, (b) ideal flat wheel, (c) rounded flat wheel. 
	\mbox{\protect\includegraphics{line1} Model,} 0.5~m/s, 
	\mbox{\protect\includegraphics{line5} experiment,} 0.5~m/s, 
	\mbox{\protect\includegraphics{line2} model,} 0.9~m/s, 
	\mbox{\protect\includegraphics{line6} experiment,} 0.9~m/s, 
	\mbox{\protect\includegraphics{line3} model,} 1.6~m/s, 
	\mbox{\protect\includegraphics{line7} experiment,} 1.6~m/s.}
	\label{fig:Ln_compSpeed}
\end{figure}

\cref{fig:Ln_compSpeed} shows the influence of changing the trolley speed in both the model and the experimental results. The tendency of the sound level to increase with increasing speed is present in both model and experimental results here. The degree of change in sound level also appears to be similar between experimental and model results, though the absolute levels remain reduced (for the smooth and rounded flat wheel) and elevated (for the ideal wheel) from the model. Interestingly, there is no measurable frequency shift with changing trolley speed. In theory the frequency should increase with speed, as the excitation is directly related to the roughness wavelengths excited in the contact area (being a function of the trolley speed). However, this effect is either not present here, or so small that it cannot be observed at the measured speeds.

\begin{figure}[!htbp]
	\centering
	\subfloat[][]{\includegraphics{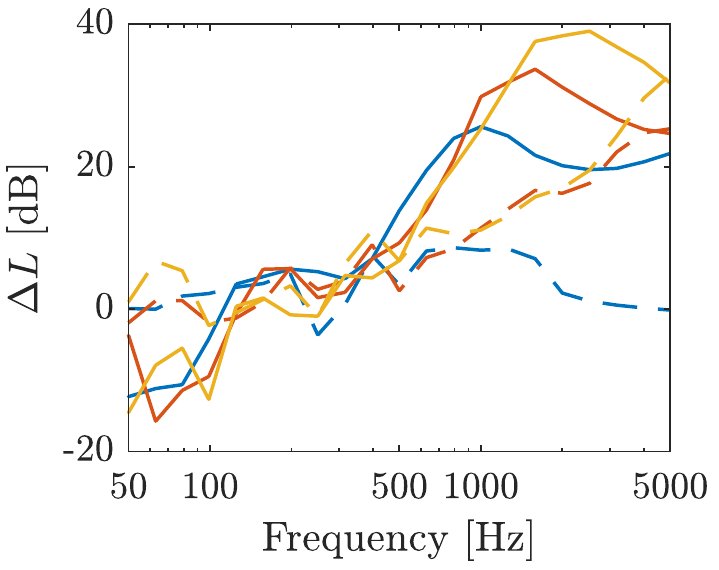}}\
	\subfloat[][]{\includegraphics{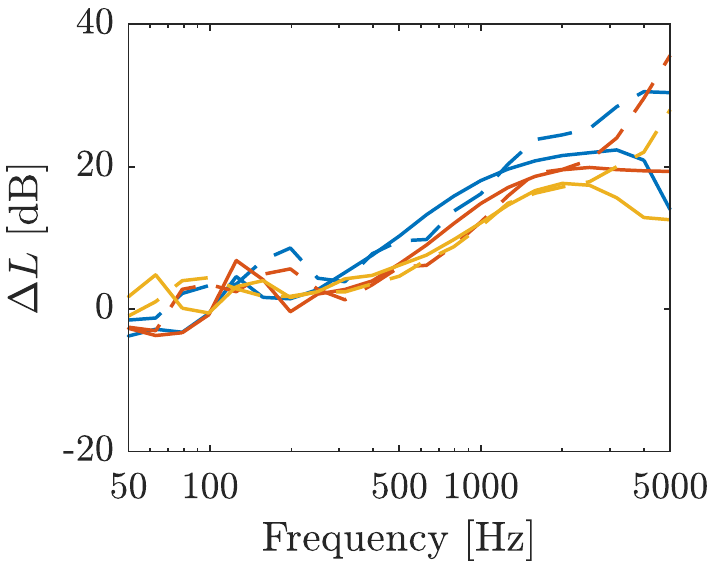}}
	\subfloat[][]{\includegraphics{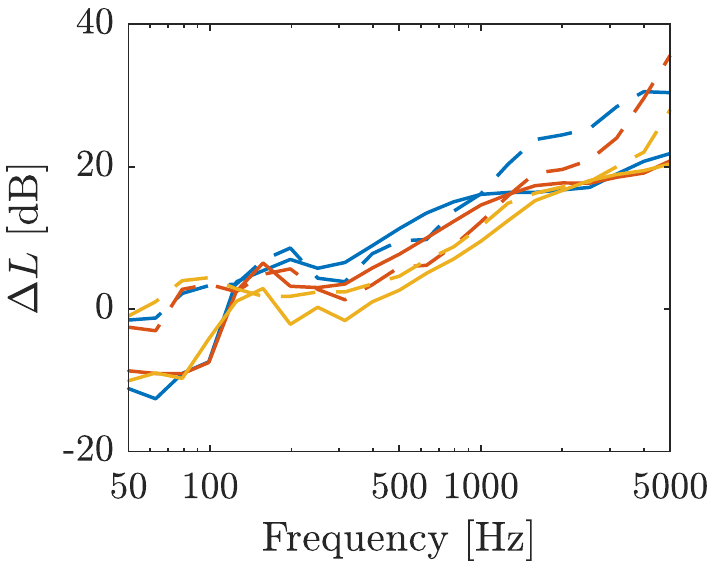}}
	\caption{Benefit from applying the rough PVC floor covering for varying trolley speed: model versus experiment. (a) Smooth wheel, (b) ideal flat wheel, (c) rounded flat wheel. 
	\mbox{\protect\includegraphics{line1} Model,} 0.5~m/s, 
	\mbox{\protect\includegraphics{line5} experiment,} 0.5~m/s, 
	\mbox{\protect\includegraphics{line2} model,} 0.9~m/s, 
	\mbox{\protect\includegraphics{line6} experiment,} 0.9~m/s, 
	\mbox{\protect\includegraphics{line3} model,} 1.6~m/s, 
	\mbox{\protect\includegraphics{line7} experiment,} 1.6~m/s.}
	\label{fig:deltaL_compSpeed}
\end{figure}

\cref{fig:deltaL_compSpeed} shows how the benefit of adding a rough floor covering changes with trolley speed for both model and experimental results. Here what was observed in the previous figure is confirmed: the model is largely capable of handling changes in the relative sound level $\Delta L$ with changes in trolley speed. The results are however still slightly more accurate for the flat wheel rather than the smooth wheel. It is interesting to note that while the trolley speed greatly affects the absolute sound level, the relative relationship between floor covering and trolley speed remains quite small. That is to say, the benefit provided by a given floor covering is largely independent of trolley speed.

\subsection{Floor admittance}
\begin{figure}[!htbp]
	\centering
    \includegraphics{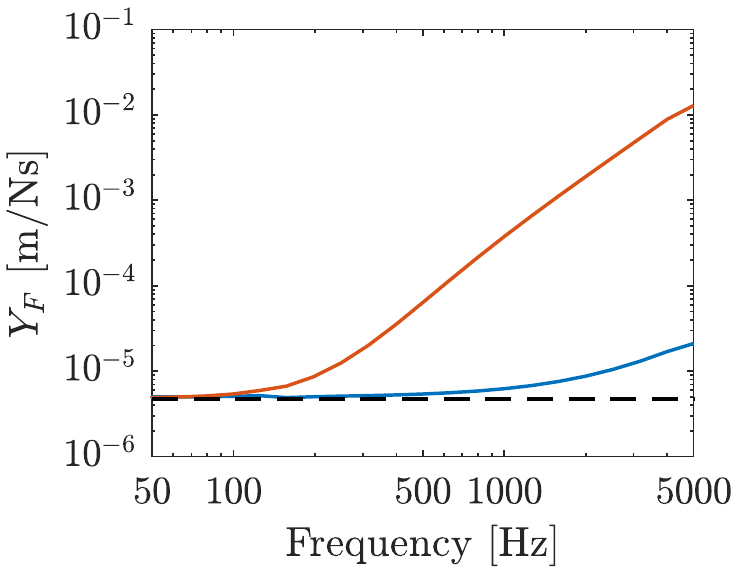}
	\caption{Unit admittance of the 100~mm concrete floor with and without PVC floor coverings. 
	\mbox{\protect\includegraphics{line1} concrete}, 
	\mbox{\protect\includegraphics{line2} concrete} + 0.6~mm PVC floor covering, 
	\mbox{\protect\includegraphics{line9} concrete} thin plate estimation.
	\cite{cremerStructureBorneSoundStructural2005}.}
	\label{fig:floor_admittance}
\end{figure}

The propagation model presented in \cref{sec:model} uses the floor admittance $Y_F$ to calculate the sound level in the reception room via the TMM. As stated previously, in practice this is done by using a unit injected force of 1~N to obtain the floor impedance before multiplying by the blocked force calculated in the rolling model to obtain the true injected force. \cref{fig:floor_admittance} shows the unit admittance of a 100~mm thick concrete floor. In addition, the unit admittance of the multi-layer floor system formed by placing a 6~mm PVC floor covering on top of the concrete floor is also shown. Finally, the thin plate approximation of the 100~mm concrete slab is shown for comparison \cite{cremerStructureBorneSoundStructural2005}. This can been seen as a flat curve which does not change with frequency. As the unit floor impedance does not depend on the surface roughness, there is no difference here between the smooth and rough PVC. These admittances were calculated using AlphaCell: a software based on the TMM which predicts the vibro-acoustic response of multi-layer systems \cite{jaouenAlphaCell2020}.

The admittance rises significantly above 100~Hz when the PVC floor covering is added, as seen in the top curve in the figure. Here less energy is transferred through the system: reducing the radiated sound level, as was shown in \cref{sec:modelLn}, and as one would logically expect with the addition of a softer floor covering. While using the admittance of a thin plate is largely appropriate for the bare concrete slab (under certain frequency and thickness conditions), the addition of a softer floor covering immediately causes the admittance to deviate strongly from the thin plate approximation for all but the lowest of frequencies. Thus, it would not be an appropriate representation in the case of a multi-layer system.

\subsection{Investigation of the contact force} \label{sec:investigation_contact_force}
Due to the nature of rolling contact, the force between the wheel and floor cannot be measured directly. Model results for the contact force due to small scale roughness, wheel flats, and floor joints are nevertheless presented below. They are discussed in the context of the characteristics they are expected to exhibit based on the known physical phenomena occurring in each scenario.

As a wheel flat moves through the vicinity of contact, the longitudinal position of the contact area relative to the wheel center will shift: first backwards as the wheel flat enters the vicinity of contact, and then forwards as the wheel flat exits the vicinity of contact. As the face of the wheel flat impacts the floor, a spike in the contact force will occur, and the position of the wheel center will drop momentarily. All of these phenomena may be visualized via the rolling model.

\begin{figure}[ht!]
	\centering
	\hspace{-2.15cm}\subfloat[Wheel center vertical displacement]{\includegraphics{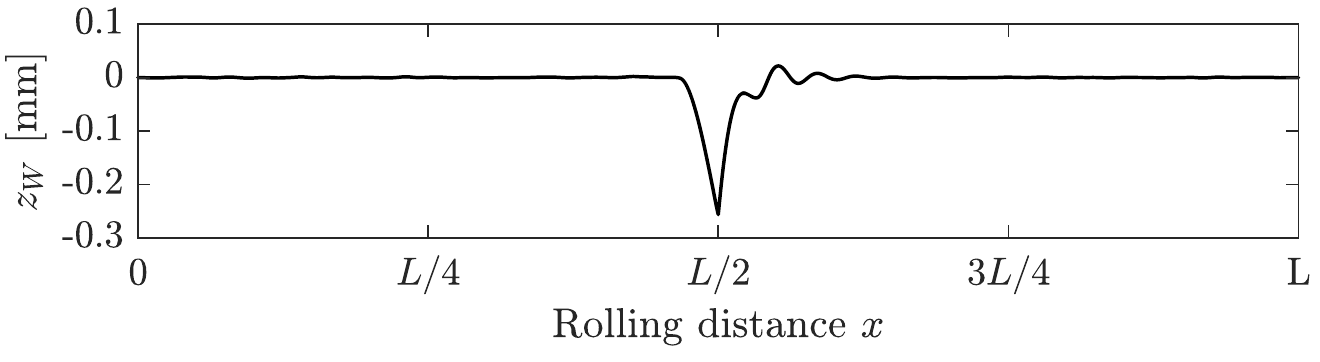}} \\
	\vspace{-0.5cm}\hspace{-2.15cm}\subfloat[Total force]{\includegraphics{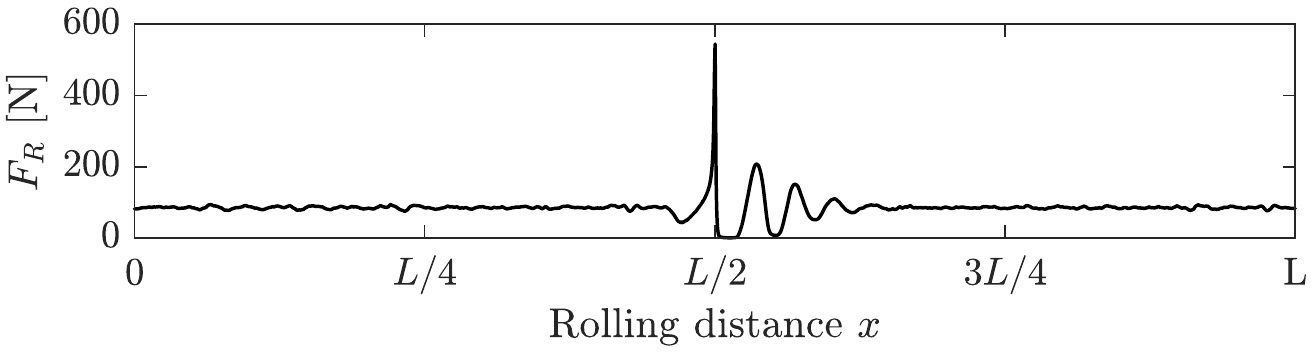}} \\
	\vspace{-0.4cm}\subfloat[][Longitudinal force distribution]{\includegraphics{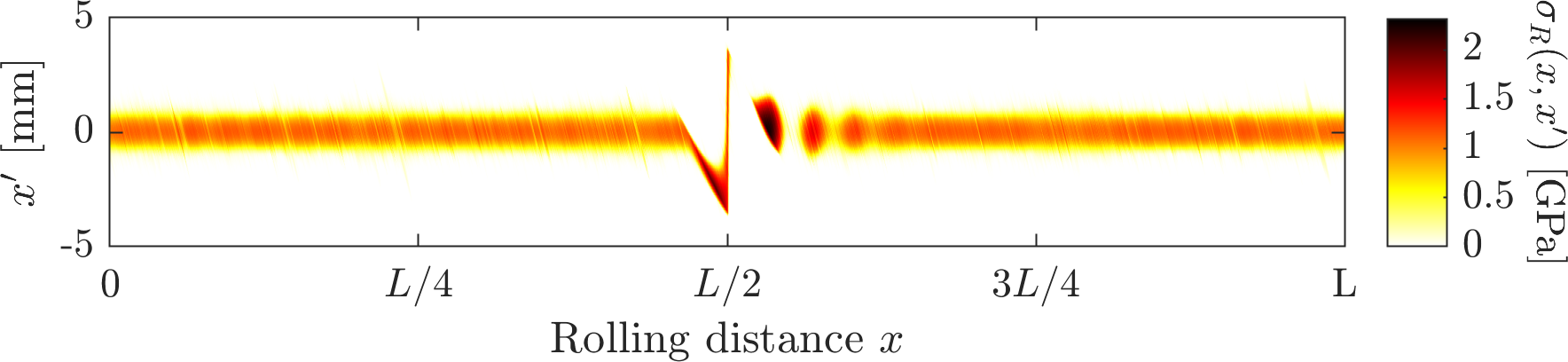}}
	\caption{Response of a wheel rolling through a single ideal wheel flat}
	\label{fig:contact_force_fConcrete_wFlatIdeal}
\end{figure}

\cref{fig:contact_force_fConcrete_wFlatIdeal} shows the vertical displacement of the wheel, contact force, and contact force distribution as a cylindrical wheel rolls through an ideal flat spot on a smooth concrete floor. Here the force distribution plot is showing slices of the contact stress profile for each discretized moment along the rolling distance $x$. Recall that $x'$ is the local coordinate system in the vicinity of contact, centered at the wheel center position. The magnitude of the contact stress appears very high because the size of a given discretized cell $\Delta x\times\Delta y$ is very small ($\Delta x = \Delta y = 55$~\textmu m). The convergence of the spatial resolution in the model has been checked.

For this simulation, the wheel ($r_x = 50$~mm) was given a single ideal flat spot ($h_W = 0.5$~mm) located at $\theta = \pi$, and the rolling distance was set to the circumference of the wheel. As such, the face of the wheel flat impacts the floor at half the rolling distance, which can be clearly seen in all three plots. Recall that the model reduces the flat spot height in order to correctly map to the reduced wheel radius $r'_x$, here being a factor of two for a cylindrical wheel. The wheel center drops a height approximately equal to the reduced wheel flat depth $h'_W$ (with small deviations from that value being due to the small scale roughness), and then returns to its previous position as the wheel flat leaves the vicinity of contact. At the moment of impact, a force is generated which is $6.4\times$ the static load (this factor is not universal and changes depending on the influencing factors such as the wheel flat geometry and trolley speed).

\begin{figure}[ht!]
	\centering
	\hspace{-2.15cm}\subfloat[Wheel center vertical displacement]{\includegraphics{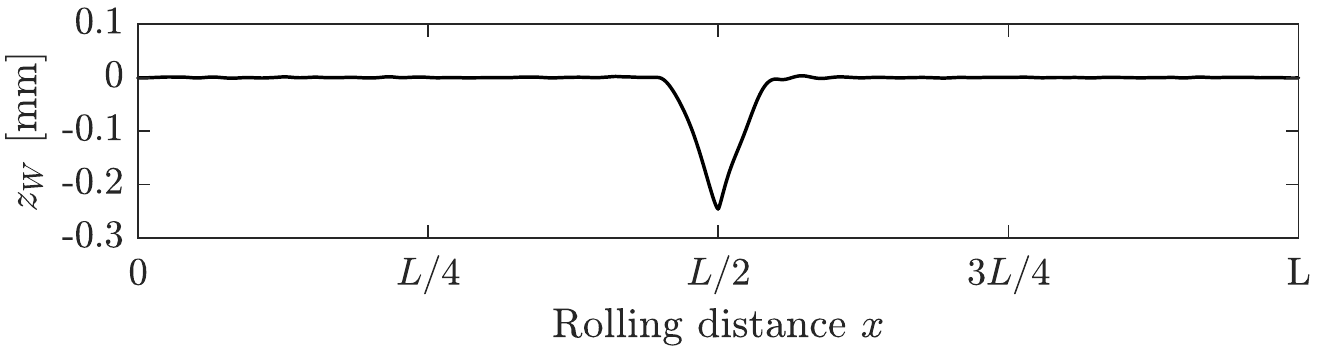}} \\
	\vspace{-0.5cm}\hspace{-2.15cm}\subfloat[Total force]{\includegraphics{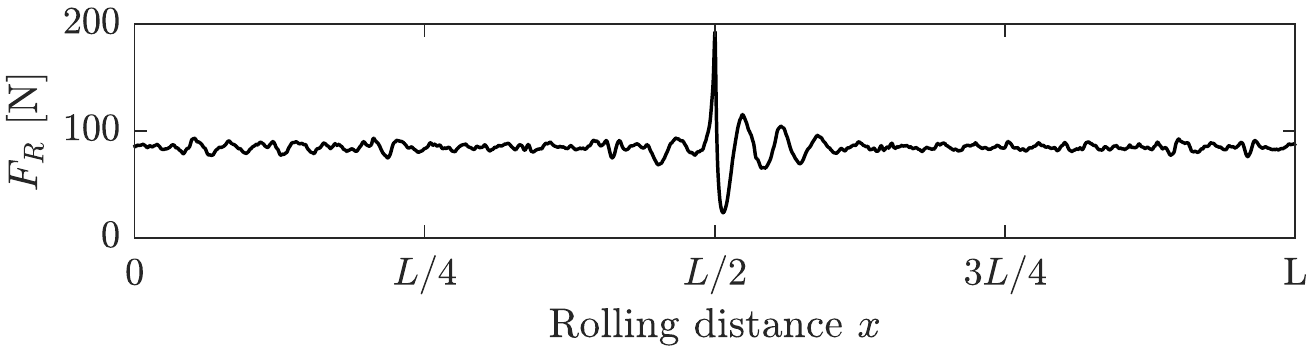}} \\
	\vspace{-0.4cm}\subfloat[][Longitudinal force distribution]{\includegraphics{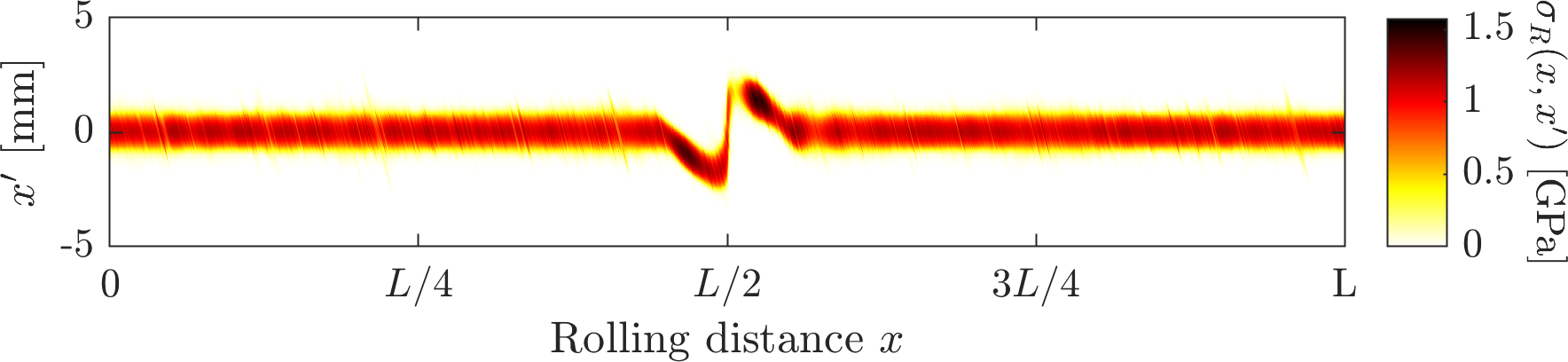}}
	\caption{Response of a wheel rolling through a single rounded wheel flat}
	\label{fig:contact_force_fConcrete_wFlatRounded}
\end{figure}

\cref{fig:contact_force_fConcrete_wFlatRounded} shows the vertical displacement of the wheel, contact force, and contact force distribution as a cylindrical wheel rolls through a rounded wheel flat on a smooth concrete floor. The same parameters were used as before, with only the type of wheel flat being changed from ideal to rounded flat. The impact force can once again be seen, though its magnitude is noticeably lower than that of the ideal wheel flat, at only $2.3\times$ the static load. The contact force distribution also exhibits a more ``smeared'' appearance, as the hard corner of the ideal wheel flat is no longer present.

\begin{figure}[h!]
	\centering
	\hspace{-2.15cm}\subfloat[][Wheel center vertical displacement]{\includegraphics{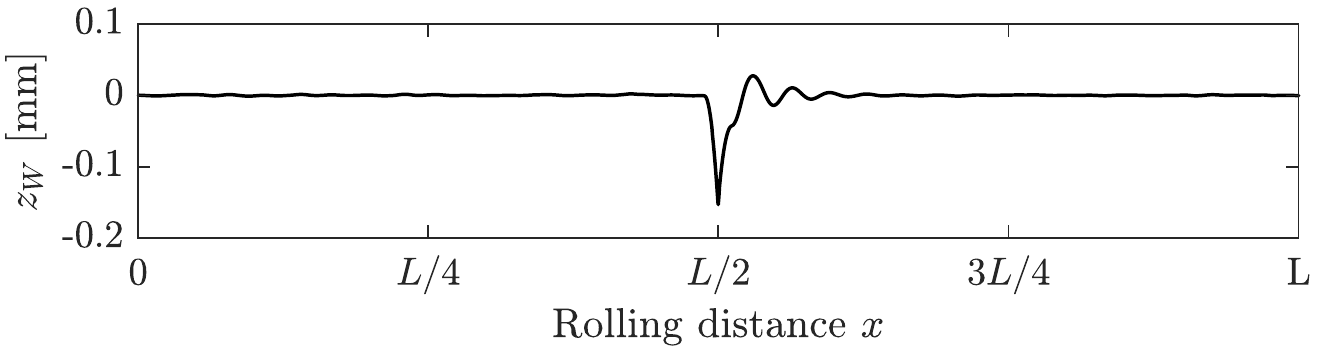}} \\
	\vspace{-0.5cm}\hspace{-2.15cm}\subfloat[][Total force]{\includegraphics{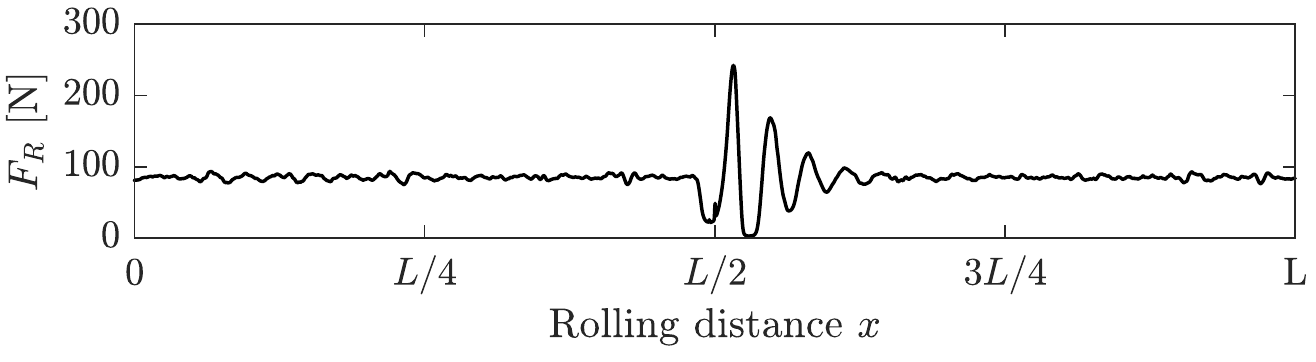}} \\
	\vspace{-0.4cm}\subfloat[][Longitudinal force distribution]{\includegraphics{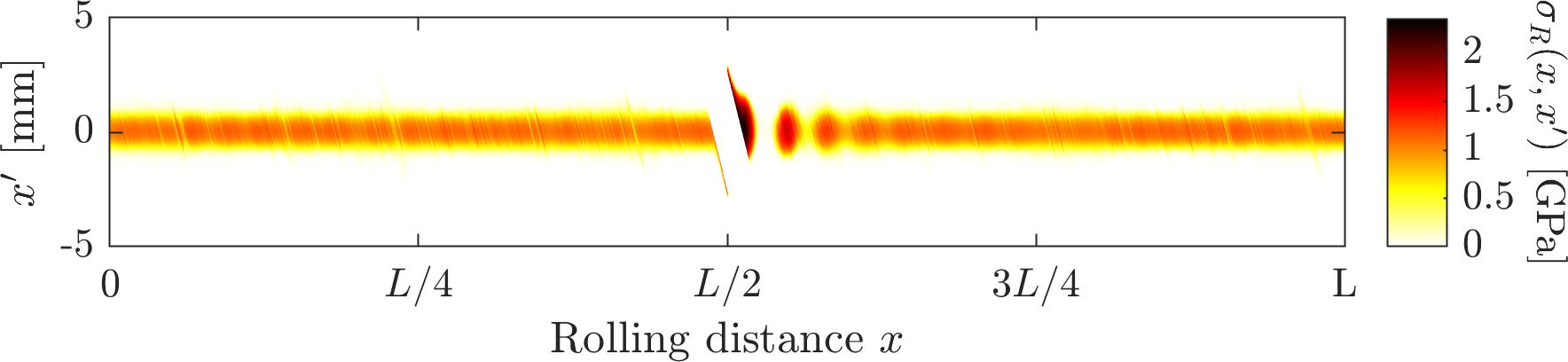}}
	\caption{Response of a wheel rolling through a single floor joint}
	\label{fig:contact_force_fConcreteJointed_wSmooth}
\end{figure}

When the wheel flat is replaced by a floor joint, a similar behavior occurs. The wheel center drops until the leading face of the wheel strikes the far side of the floor joint. \cref{fig:contact_force_fConcreteJointed_wSmooth} shows the vertical displacement of the wheel, contact force, and contact force distribution for a smooth wheel rolling over a floor joint. The wheel here ($r_x = 50$~mm) encountered a single floor joint ($l_F = 1$~cm) located half the rolling distance (which was again set to the circumference of the wheel). The drop in wheel center is once again visible as the wheel traverses the floor joint. The fact that the area of contact jumps from behind to in front of the wheel center can be seen in the contact force distribution. This differs from the behavior of a wheel flat, as the wheel does not contact the bottom of the floor joint. There is, however, no large spike in the contact force at the moment of impact with the far side of the floor joint. The spike instead comes slightly afterward at $2.9\times$ the static load, as the wheel moves up and out of the far side of the floor joint. A very small spike can be seen at the moment of impact (at $L/2$), but its magnitude is not much larger than that which would be caused by a large roughness asperity.

In all the total force plots from \cref{fig:contact_force_fConcrete_wFlatIdeal,fig:contact_force_fConcrete_wFlatRounded,fig:contact_force_fConcreteJointed_wSmooth}, the contact force due to the small scale roughness (i.e. out of the vicinity of the discrete irregularity) modulates about the static load $Q = 85$~N, indicating that the magnitude of the contact force estimated by the model is correct.

\subsection{Contact force spectrum}
\begin{figure}[ht!]
	\centering
	\subfloat[][Smooth wheel]{\includegraphics{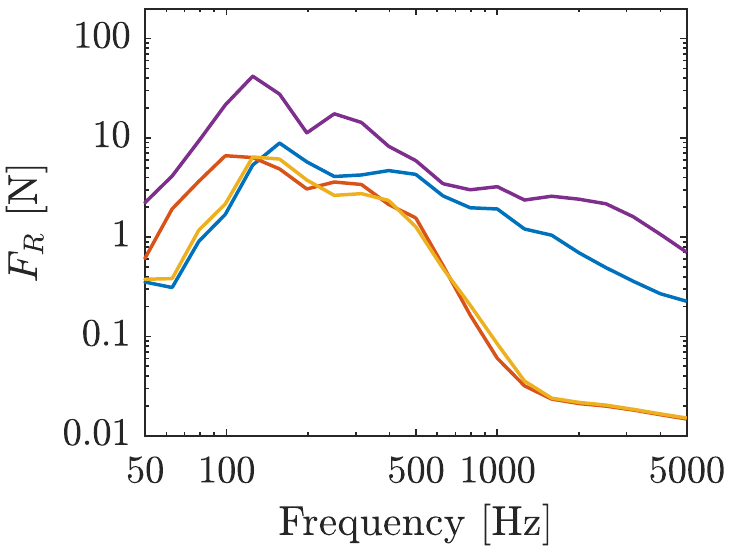}}\\
	\subfloat[][Ideal flat wheel]{\includegraphics{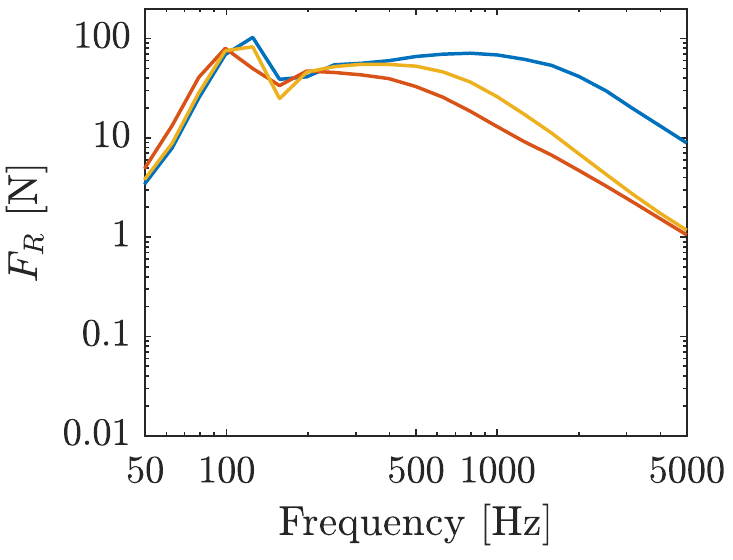}}
	\subfloat[][Rounded flat wheel]{\includegraphics{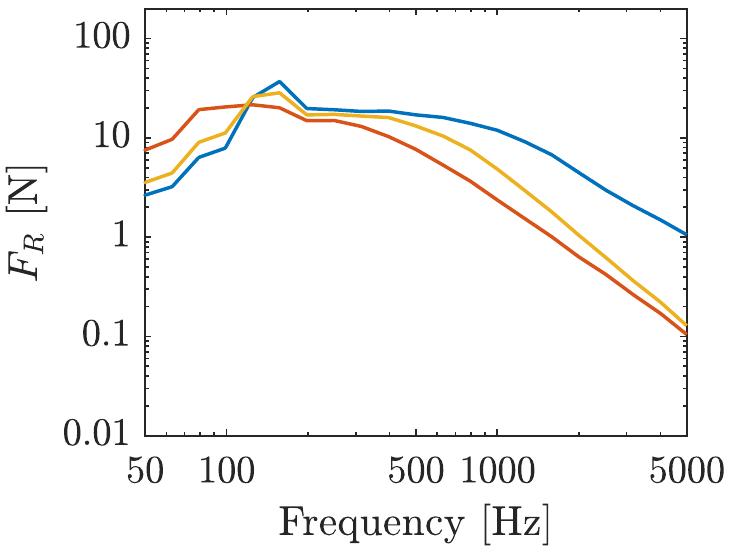}}
	\caption{Contact force spectra calculated by the rolling noise model. 
	\mbox{\protect\includegraphics{line1} Smooth} concrete floor, 
	\mbox{\protect\includegraphics{line2} rough} PVC floor, 
	\mbox{\protect\includegraphics{line3} smooth} PVC floor, 
	\mbox{\protect\includegraphics{line4} jointed} concrete floor.}
	\label{fig:SFr}
\end{figure}

\cref{fig:SFr} shows the contact force spectra $F_R(f)$ for each of the three wheels. The jointed concrete yields the highest magnitude force, followed by the smooth concrete, and finally two PVC floor coverings. In the case of the smooth wheel, the two PVC floor coverings have very similar force spectra. In the case of the flat wheel, the smooth PVC has a slightly higher force spectrum in the mid frequency range than the rough PVC. This matches the trend seen in the normalized sound pressure level data. The shape of the curves are similar to those in \cref{fig:Ln_compFloor,fig:Ln_compMass,fig:Ln_compSpeed}. This is to be expected above the coincidence frequency of the floor (250 Hz), as the radiation efficiency of the floor is moderately uniform in this region. 

\subsection{Wheel impedance}
\begin{figure}[ht!]
	\centering
	\subfloat[][Smooth wheel]{\includegraphics{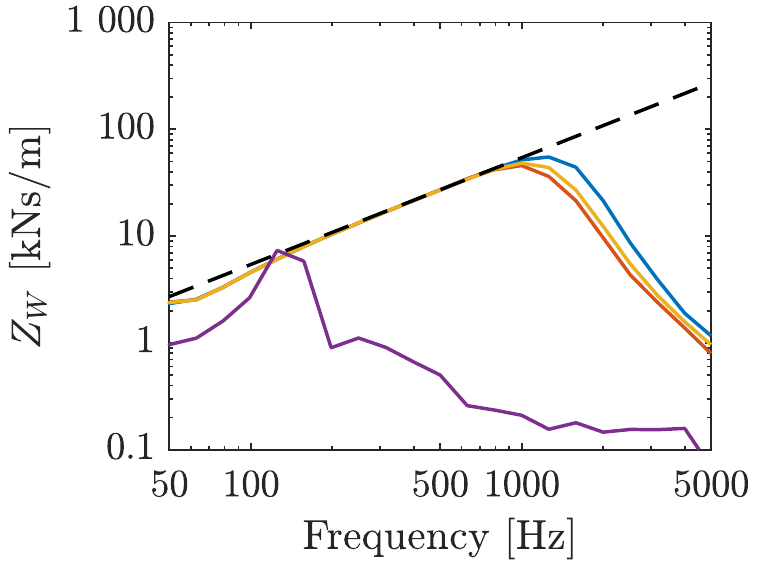}}\\
	\subfloat[][Ideal flat wheel]{\includegraphics{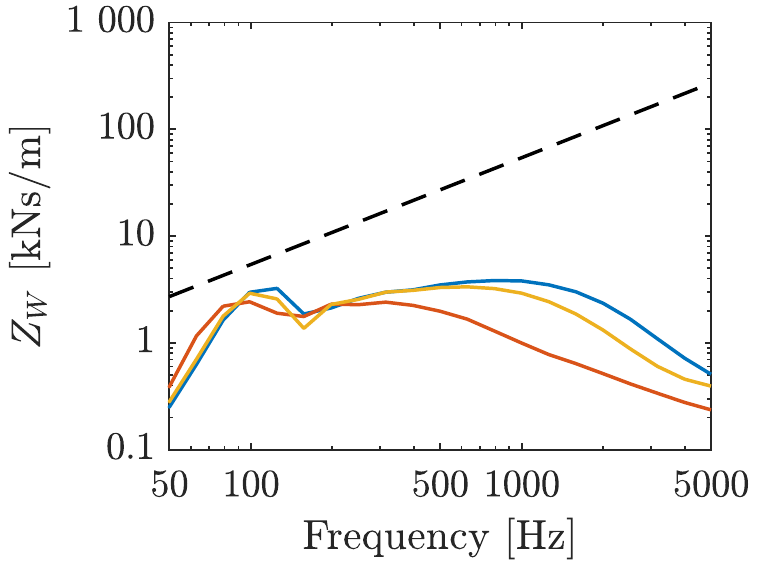}}
	\subfloat[][Rounded flat wheel]{\includegraphics{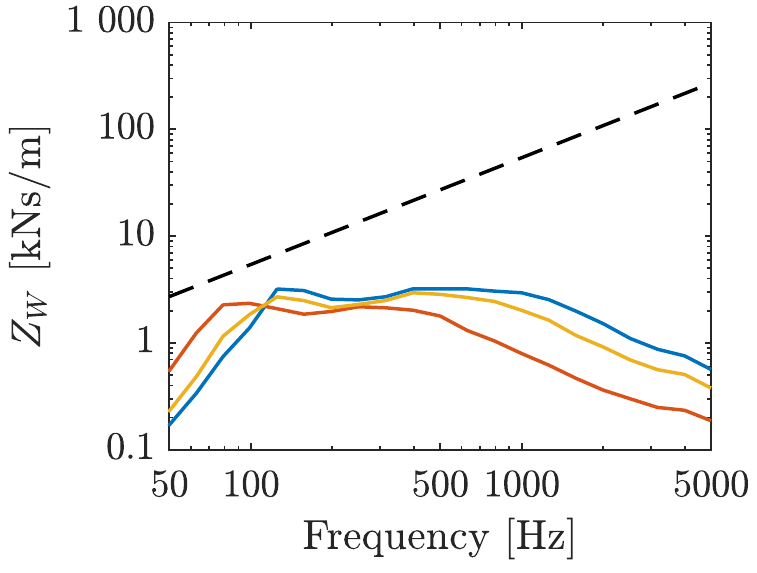}}
	\caption{Wheel impedance spectra calculated by the rolling noise model. 
	\mbox{\protect\includegraphics{line1} Smooth} concrete floor, 
	\mbox{\protect\includegraphics{line2} rough} PVC floor, 
	\mbox{\protect\includegraphics{line3} smooth} PVC floor, 
	\mbox{\protect\includegraphics{line4} jointed} concrete floor, 
	\mbox{\protect\includegraphics{line9} simple}  mass: $i\omega M$.}
	\label{fig:SZw}
\end{figure}

\cref{fig:SZw} shows the wheel impedance spectra $Z_W(f)$ for each of the three wheels. The impedance is calculated by dividing the contact force by the wheel vertical velocity. For comparison, the impedance of a simple rigid  mass ($i\omega M$) is also shown (e.g. as is the case for the hammer of a tapping machine). In the absence of discrete irregularities, the wheel impedance follows that of a simple mass, deviating only above 1~000~Hz where damping and elastic effects play a greater role. The curve of the simple mass continues upward from there. The impedance is significantly lower for the flat wheels and jointed floor. Similar to the $L_n$ curves, the wheel impedance for the smooth wheel on the jointed concrete floor is more similar to that of the rounded wheel on smooth concrete than that of the smooth wheel on smooth concrete. While using the impedance of a simple mass would be mostly appropriate in the absence of discrete irregularities, the unique wheel impedance calculated by the model is necessary when they are to be included. In all cases, the wheel impedance on smooth concrete is slightly higher than smooth PVC, which is in turn slightly higher than rough PVC. Because the wheel response (e.g. vertical velocity) depends on the type of floor it is rolling on, its impedance is dependent on the type of floor as well.

\section{Discussion}
Due to the nature of rolling noise, with experimental results showing maxima in the range of 200--500 Hz, as well as the ease with which it propagates throughout multi-story building structures, developing a model that can accurately predict it has the opportunity to lead to the development of flooring materials which are effective at attenuating multiple sources of annoyance, not just impact noise. The potential for such a model to be a useful tool in building design (and acoustic material research) lies in its ability to quickly analyze multiple combinations of wheels and floors to identify trends that may exist.

The primary objective of this work was to develop a rolling noise model that can capture the physical phenomena present in indoor rolling contact, as well as predict the sound level benefit of adding a floor covering to a given floor. The model which was developed has been specifically adapted for the characteristics of rolling in multi-story buildings, such as highly elastic wheels, multi-layer floors, and sound propagation via direct vertical structure-borne transfer paths. This is a domain which has hereto been left unexplored. As such, this work serves as a first look into the problem of indoor rolling noise. This work serves as a platform which may spur further exploration into the field of indoor rolling noise.

An important observation to make when comparing the two plots in \cref{fig:Ln_compFloor} is that, while the addition of a PVC floor covering aids in reducing the overall noise level for a given wheel, the flat wheel on PVC is still louder than the smooth wheel on bare concrete. This validates the necessity for a model which is capable of accounting for discrete irregularities, as their presence can quickly eliminate any benefit that a quieter material (whether for the floor or wheel) would provide. The smooth wheel does not always provide a sufficient SNR for precise measurements, particularly in the presence of a soft floor covering.

Using a model such as the one presented here, the acoustic performance of various floor systems may be estimated in order to identify which ones yield the greatest benefit for rolling noise. This is a technique which already exists for other sources of indoor structure-borne noise, such as the tapping machine. However, because impact noise is starkly different than rolling noise, floor systems which have been developed for reducing impact noise will not necessarily exhibit high performance with rolling noise as well. This model may serve as a compliment to the existing indoor structure-borne noise modeling solutions: aiding in the development of building materials which have better acoustic performance for a wider range of sound sources.

As stated previously, the propagation model accounts for the finite size of the floor, but not its exact modal behavior. The model is capable of using a measured floor mobility rather than the theoretical admittance calculated with the TMM. Doing so, however, reduces the scope of the model's applicability, as it can only be used in this state for floors for which experimental data has been measured.


There exists room for improvement in the model regarding the influence of the added load on the trolley. The inverse relationship between increasing load and sound level observed in the experimental data at low frequencies is not seen in the model results, and thus could stand to be improved.

As is the case with any numerical model, the results are only as good as the inputs. In addition to the roughness profiles, experimental data regarding the elastic properties of the floor (in particular, the dynamic loss factor) are necessary for this model. It is known that the damping of the floor has a large effect on the transmitted structure-borne noise when being excited by a mechanical source \cite{chevillotteAnalysisExcitationsWavenumber2015}. While extensive measures were taken to obtain highly accurate roughness profiles, improved measurement of the wheel and floor material properties may provide greater accuracy in the overall model results.

The test trolley used was developed to produce a rolling sound which was as regular as possible: free of extraneous signatures such as rattling or other trolley vibrations. Just as there exists a standard tapping machine for normalizing the measurement process of impact noise across various test locations, the possibility of a standard rolling device also exists. The development of such a device could aid in the furthering the field of indoor rolling noise research, as it would allow a congruent means of comparing rolling noise measurements gathered in different scenarios. For a further discussion on this topic, see \cite{edwardsRecommendationStandardRolling2020}. For more in depth exploration of the experimental testing and model development presented in this paper, see \cite{edwardsPredictionControlRolling2020}.
 
\section*{Conclusion}
This paper presents a time-domain model for predicting rolling noise in buildings. The model is able to predict the radiated sound pressure level, as well as the relative benefit provided by a floor covering, in response to indoor rolling noise in a multi-story building environment. For the overall performance of a building, the focus is on the radiated sound pressure level. When developing a new flooring system, the focus is on the relative sound benefit. The model accurately represents the physical phenomena present in indoor rolling noise, accounting for influencing factors such as the roughness of the floor \& wheel, the elastic parameters of the floor \& wheel, the speed of the trolley, and the load on the trolley. The model is also able to account for discrete irregularities, such as wheel flat spots and floor joints. A model such as this one may be used as a supplement to the existing models for floor impact noise, in order to test different wheel/floor material combinations to see which ones result in the greatest reduction in noise. It may also be used to identify trends in order to guide the development of new vibro-acoustic materials which are effective at reducing both impact noise and rolling noise.

\section*{Acknowledgments}
This work was done as part of Acoutect: an innovative training network composed of 5 academic and 7 non-academic participants. This consortium comprises various disciplines and sectors within building acoustics and beyond, promoting intersectoral, interdisciplinary and innovative training and mobility of the researchers within the project. This project has received funding from the European Union’s Horizon 2020 research and innovation program under grant agreement No 721536. This work was also  performed within the framework of the Labex CeLyA of Université de Lyon, operated by the French National Research Agency (ANR-10-LABX-0060/ANR-11-IDEX-0007).

\bibliographystyle{ieeetr}
\bibliography{RollingModel}

\end{document}